\documentclass[12pt]{article}

\usepackage{epsfig}
\usepackage{amsmath,amssymb}
\usepackage{bm}
\usepackage{bbm}
\usepackage{color}
\usepackage{slashed}

\newcommand{\sign}{{\rm sign}}

\newcommand{\lsim}
{\;\raisebox{-.3em}{$\stackrel{\displaystyle <}{\sim}$}\;}

\newcommand{\SigSlash}{\Sigma \!\!\! /}

\begin{document}
\thispagestyle{empty}

\begin{flushright}
{
\small
TUM-HEP-880-13\\
TTK-13-07\\
ANL-HEP-PR-13-19
}
\end{flushright}

\vspace{0.4cm}
\begin{center}
\Large\bf\boldmath
Scattering Rates For Leptogenesis:\\
Damping of Lepton Flavour Coherence and
Production of Singlet Neutrinos
\unboldmath
\end{center}

\vspace{0.4cm}

\begin{center}
{\textbf{Bjorn Garbrecht},$^{1,2}$ \textbf{Frank Glowna},$^{1,2}$ and \textbf{Pedro Schwaller}\,$^{3,4}$}\\
\vskip0.3cm
$^1${\it
Physik Department T70, James-Franck-Strasse, \\Technische Universit\"at M\"unchen, 85748 Garching, Germany}

\vskip0.3cm
$^2${\it Institut f\"ur Theoretische Teilchenphysik und Kosmologie,\\
RWTH Aachen University,
52056 Aachen, Germany}\\
\vskip.3cm

$^3${\it HEP Division, Argonne National Laboratory, 
\\Argonne, IL 60439, U.S.A.}
\vskip.3cm

$^4${\it Department of Physics, University of Illinois at Chicago,
\\Chicago, IL 60607, U.S.A.}
\end{center}

\vspace{2cm}
\begin{abstract}
Using the Closed-Time-Path approach, we perform
a systematic leading order calculation of the relaxation rate
of flavour correlations of left-handed Standard Model leptons. This quantity is of pivotal relevance for flavoured
Leptogenesis in the Early Universe, and we find it to be
$5.19\times 10^{-3}T$ at $T=10^7\,{\rm GeV}$ and $4.83 \times 10^{-3}T$
at $T=10^{13} {\rm GeV}$. These values apply to the Standard Model with a Higgs-boson mass of $125\,{\rm GeV}$. The dependence of the numerical
coefficient on the temperature $T$ is due to the renormalisation group
running. The leading linear and logarithmic dependencies of the flavour relaxation rate on the gauge and top-quark couplings are extracted,
such that the results presented in this work can readily be applied to extensions of the Standard Model. We also derive the production rate
of light (compared to the temperature) sterile right-handed neutrinos,
a calculation that relies on the same methods. We confirm most details
of earlier results, but find a substantially larger contribution
from the $t$-channel exchange of fermions.
\end{abstract}
\newpage

\tableofcontents

\newpage

\section{Introduction}

The origin of the matter anti-matter asymmetry in the Universe is one of the most important open questions in Particle Physics and Cosmology. Leptogenesis~\cite{Fukugita:1986hr}\footnote{See e.g. Refs.~\cite{Buchmuller:2012eb,Blanchet:2012bk,Fong:2013wr} for recent reviews.} is among the most plausible possible mechanisms for generating the observed baryon asymmetry dynamically in the Early Universe. The recent discovery of a Standard Model (SM) like Higgs boson provides further support for this mechanism, where the asymmetry is generated by out-of-equilibrium decays of heavy sterile
right-handed neutrinos into a Higgs boson and a  lepton. 
 
In recent years, substantial progress was made in the theoretical description of Leptogenesis using Non-Equilibrium Quantum Field Theory in the Closed Time Path (CTP) formalism~\cite{Schwinger:1960qe,Keldysh:1964ud,Calzetta:1986cq}. Particular attention was given to the derivation of quantum evolution equations for distribution functions~\cite{Buchmuller:2000nd,De Simone:2007rw,Cirigliano:2009yt,flbg,Anisimov:2010dk},  the calculation of thermal corrections to the charge-parity ($CP$) asymmetry~\cite{Covi:1997dr,Giudice:2003jh,Garny:2009rv,Garny:2009qn,Anisimov:2010aq,Garny:2010nj,kblg}, and related questions~\cite{Garbrecht:2010sz,Garbrecht:2011aw,Garny:2011hg,Garbrecht:2012qv,Drewes:2012ma,Garbrecht:2012pq,Frossard:2012pc,Millington:2012pf}.

Interaction rates mediated by Higgs-Yukawa couplings are crucial for Leptogenesis predictions. The production and decay rates of the lightest right-handed neutrino determine the strength of the washout, and they are mediated by their Yukawa couplings
$Y$ to SM lepton doublets and Higgs bosons, while the lepton flavour equilibration rate determines the temperature scale where flavour effects become relevant for calculations of the asymmetry~\cite{Endoh:2003mz,Abada:2006fw,Nardi:2006fx}, and it is mediated by the SM lepton Yukawa
couplings $h$.

At low temperatures, both processes [right-handed neutrino (inverse) decays and
flavour equilibration] are described at leading order (LO) through $1\leftrightarrow2$ processes, namely decays and inverse decays, involving only the Higgs boson and left- and right-handed leptons and sterile neutrinos. At high temperatures, both rates are phase space suppressed. Therefore a complete LO calculation of these crucial interactions must include processes where additional gauge bosons are emitted or absorbed or where
an intermediate Higgs boson decays into a pair of top quarks. For
right-handed neutrino (inverse) decays, the LO analysis pertinent to
high temperatures has first been performed in Refs.~\cite{Anisimov:2010gy,Besak:2012qm}.
In some earlier and also more recent papers~\cite{Frossard:2012pc,Kiessig:2010pr,Kiessig:2011fw,Kiessig:2011ga},
the inclusion of the gauge interactions
is restricted to the effect of the modified thermal dispersion relations (thermal masses),
but these approaches should only partly capture the LO effects.
Closely related to the problem of right-handed neutrino production and
the relaxation of the flavour correlations of active leptons
is the production of photons in a quark-gluon plasma, which
is calculated {\it e.g.} in Refs.~\cite{Arnold:2000dr,Arnold:2001ms}.

Gauge boson corrections in the low temperature regime have recently received some attention in the works~\cite{Salvio:2011sf,Laine:2011pq}, where a  non-relativistic
analysis including leading relativistic corrections has been performed. When
the right-handed neutrino mass cannot be neglected, the tree-level scattering
amplitudes exhibit infrared (IR) divergences, and the familiar cancellation of these with
contributions from virtual diagrams has to be generalised to a finite-temperature
environment. It turns out that the cancellation of IR divergences in the non-relativistic
regime can be generalised to fully relativistic processes as
well~\cite{Garbrecht:2013gd}.

The main topic of this paper is the calculation of Higgs Yukawa mediated interaction rates in a high temperature background. We have performed this
calculation using methods based on the two-particle-irreducible (2PI) CTP formulation of Non-Equilibrium Quantum Field Theory, within the framework that has been
applied to calculate $CP$-violating rates in Refs.~\cite{flbg,kblg}. A main advantage of the 2PI approach is that it directly gives
rise to a derivation of
the appropriate description of the physical screening
that regulates the $t$-channel divergences that occur in certain tree-level scattering diagrams. The screening is implemented by the use
of a resummed propagator, which in contrast appears as an ad hoc
prescription within an approach based on the calculation of
scattering matrix elements.

Presently,
our main phenomenological interest is to perform a LO calculation
of the damping rate of lepton flavour coherence, that is of importance for
flavoured Leptogenesis. This flavour damping rate can be used as an input
to the systematic analysis of flavour decoherence in Leptogenesis, that
has been presented in Ref.~\cite{flbg}. Having obtained the various diagrammatic
contributions to flavour damping, it is a simple matter to extract the
(inverse) decay rates of right-handed neutrinos in the high temperature regime.
We thereby confirm most details of the results
presented in Refs.~\cite{Anisimov:2010gy,Besak:2012qm}, but find a substantially
larger contribution from $t$-channel fermion exchange.
The latter discrepancy should be due
to a different technical implementation of the extraction of
the coefficients of the logarithmically
enhanced (in squares of the gauge coupling constant) and the standard perturbative
contributions. In order to test our method, we have therefore
compared its result to a fit to the numerically performed integral and found
it to agree within the expected accuracy of the present approximations.

The remainder of this paper is organised as follows: In Section~\ref{sec:setup}, we introduce the formalism and explain how the relevant interaction rates are obtained in this setup. Contributions to the interaction rates from self-energy type and vertex type diagrams are calculated in Sections~\ref{sec:self} and~\ref{sec:vertex} respectively, while collinearly enhanced $1\to 2$ processes are calculated in Section~\ref{sec:onetwo}. The final numerical results and implications for phenomenology are presented in Section~\ref{sec:pheno}, before we conclude in Section~\ref{sec:conclusions}. 

%
%

\section{Setup}
\label{sec:setup}

\subsection{Goal of the Calculation}

The interactions that lead to the production of right-handed neutrinos
$N_i$ are the Yukawa couplings $Y_{ia}$. The lepton asymmetry can be
partly transferred from left-handed lepton doublets $\ell_b$ to right
handed active leptons ${\rm R}_a$ through the Yukawa couplings $h_{ab}$.
These also lead to the decoherence of off-diagonal correlations between
the lepton flavours. Explicitly, these
Yukawa interactions are given by the
Lagrangian terms
\begin{align}
-Y_{ia}\bar \psi_{N i} \tilde \phi^\dagger P_{\rm L}\psi_{\ell a}
-h_{ab}\phi^\dagger \bar\psi_{{\rm R}a} P_{\rm L} \psi_{\ell b}
+ {\rm c.c.}\,,
\end{align}
where ${\rm c.c.}$ stands for complex conjugation. Four-component spinors
are denoted by $\psi$, and the subscripts indicate the fields that they are
associated with. While a formulation in terms of Weyl spinors
is also possible and perhaps more appropriate, we choose here four component
spinors in order to make use of the standard identities for Dirac matrices
for the simplification of the spinor algebra. The indices of ${\rm SU}(2)_{\rm L}$
are suppressed, the contraction of these between the fields $\phi$ and $\ell$
is implied, and $\tilde \phi=(\epsilon \phi)^\dagger$, where $\epsilon$
is the antisymmetric rank-two ${\rm SU}(2)_{\rm L}$ tensor.

The goal of this paper is to calculate interaction rates mediated by Yukawa couplings in a finite temperature, finite density environment like the Early Universe. Tree level $1\to 2$ processes that are proportional to $h^2$ are kinematically
suppressed in the Early Universe, since all involved particles are massless ({\it i.e.} masses much smaller than the temperature)
at temperatures $T$
above the Electroweak phase transition. Similarly, when $T\gg M_N$,
where $M_N$ is the mass of the right-handed neutrino,
the right-handed neutrinos can be approximated as massless, such that decays $N\to \phi\ell$ are suppressed.

With the tree-level $1\leftrightarrow 2$ channels forbidden or suppressed, the LO rates
involve the radiation of one additional gauge boson, which leads to scattering rates that are parametrically of order $g^2h^2$ or $g^2 Y^2$, respectively, where $g$ denotes a gauge coupling.
At the same order, there are also contributions from collinearly
enhanced $1\leftrightarrow 2$ processes when medium effects mediated by gauge interactions are resummed into the propagators~\cite{Anisimov:2010gy}.
Due to the sizable top-quark Yukawa coupling, a leading order calculation should
also account for the Higgs boson decaying into a top-quark pair.

For the present calculation, we assume that all external particles are massless (before
including thermal corrections), what leads to significant simplifications ({\it i.e.} the vanishing of numerous contributions from the CTP Feynman rules and the absence of soft
and collinear IR divergences) that become apparent during the course of the calculation.
Within the CTP formalism, the production and relaxation rates can be
inferred from the collision term, as explained in Section~\ref{sec:collision}.
The collision terms for $N$, $\ell$ and ${\rm R}$ encompass similar
diagrams, that are related among one another by an exchange
of the Yukawa coupling matrices and the gauge coupling constants.
In the following, we therefore derive the collision term for $\ell$, but
we factorise the results in such a way
that they can easily be employed in
order to obtain the production and destruction rates for ${\rm R}$
and $N$ as well.

\subsection{Definitions}
We perform the
calculation of the collision terms within the framework developed in
Refs.~\cite{flbg,kblg}. Here we just give the definitions that are relevant for the present calculation. 
The fermionic Wightman functions that appear in the collision term are 
\begin{subequations}\label{eqn:prop:fermion}
\begin{align}
{\rm i}S^<(k)&=-2S^{\cal A}(k) f(k)\,,
\\
{\rm i}S^>(k)&=2S^{\cal A}(k) [1-f(k)]\,,
\end{align}
\end{subequations}
where $S^{\cal A}(k)$ is the spectral function that determines the location of the quasi-particle poles and $f(k)$ is the distribution function that will be specified more precisely below. 
Tree-level propagators, which we indicate
by a superscript $(0)$, are recovered when replacing the spectral function $S^{\cal A}$ with a delta distribution: 
\begin{align}
\label{spec:fun:fermi:tree}
S^{(0){\cal A}}(k)=\pi P_X(\slashed k + m)\delta(k^2-m^2){\rm sign}(k^0)\,.
\end{align}
In case of massless, chiral fermions, either $P_X=P_{{\rm L},{\rm R}}$
({\it i.e.} $P_{\rm L}$ for $\ell$ and $P_{\rm R}$ for ${\rm R}$),
whereas for Dirac and four-component Majorana fermions, $P_X=\mathbbm{1}$.
The relations for the distribution functions
are
\begin{align}
\label{ffourfermi}
f(k)=\left\{
\begin{array}{l}
f(\mathbf k)\quad\textnormal{for}\quad k^0>0\\
1-\bar f(\mathbf k)\quad\textnormal{for}\quad k^0<0
\end{array}
\right.
\qquad\textnormal{(fermions)}\,,
\end{align}
where $f(\mathbf k)$ and $\bar f(\mathbf k)$ are the distribution
functions of particles and anti-particles.
In kinetic equilibrium, the distributions are of the Fermi-Dirac form
\begin{align}
f(k)=\frac{1}{{\rm e}^{\beta(k^0-\mu)}+1}\,,
\end{align}
where $\beta=1/T$.
Note that $\mu$ and $f(k)$ may be hermitian matrices
in order to take account of flavour coherence~\cite{flbg}.

To calculate the self energies, we also need the bosonic Wightman functions for the Higgs and gauge bosons. For scalars, these are
\begin{subequations}
\begin{align}
{\rm i}\Delta^<(k)&=2\Delta^{\cal A}(k) f(k)\,,
\\
{\rm i}\Delta^>(k)&=2\Delta^{\cal A}(k) [1+f(k)]\,,
\end{align}
\end{subequations}
where the tree-level spectral function for scalar particles
of mass $m$ is given by
\begin{align}
\label{specfun:scalar:tree}
\Delta^{(0){\cal A}}(k)=\pi\delta(k^2-m^2){\rm sign}(k^0)\,.
\end{align}
The distribution function $f(k)$ with the four-momentum as an argument
can be related to the particle and anti-particle
distributions $f(\mathbf k)$ and $\bar f(\mathbf k)$ as
\begin{align}
\label{ffourbose}
f(k)=\left\{
\begin{array}{l}
f(\mathbf k)\quad\textnormal{for}\quad k^0>0\\
-[1+\bar f(\mathbf k)]\quad\textnormal{for}\quad k^0<0
\end{array}
\right.
\qquad\textnormal{(bosons)}\,.
\end{align}
For distributions in kinetic equilibrium,
\begin{align}
f(k)=\frac{1}{{\rm e}^{\beta(k^0-\mu)}-1}\,.
\end{align}
Finally we use gauge-boson propagators in Feynman gauge, such that
\begin{align} {\rm i} \Delta^{<,>}_{\mu\nu} = -g_{\mu\nu} {\rm i} \Delta^{<,>}\,. \notag \end{align}
Evolution equations for the distribution functions $f(k)$ can be derived from the Dyson-Schwinger equations for the Wightman functions $S^{ab}$ and $\Delta^{ab}$. In Wigner space and to first order in gradients, the relevant equations for the the lepton doublet and for the right-handed neutrino are
\begin{align}
\label{kineqs:ell}
	&i \partial_t \gamma^0 S_{\ell}^{<,>} - \left[\mathbf{k}\cdot \mathbf{\gamma} \gamma^0 + \Sigma\!\!\!/_\ell^{H} \gamma^0, i \gamma^0 S_\ell^{<,>} \right] - \left[ i \Sigma\!\!\!/_\ell^{<,>} \gamma^0,\gamma^0 S_\ell^H \right] = -\frac{1}{2}\left( i {\cal C}_\ell + i {\cal C}_\ell^\dagger \right),\\
\label{kineqs:N}
	&i \partial_t \gamma^0  S_N^{<,>} = - i {\cal C}_{N}\,.
\end{align}
For a derivation of these equations and the further treatment of the left-hand sides we refer the reader to Refs.~\cite{flbg,kblg}. Physical processes, like the Yukawa induced scattering and decay processes that we are interested in here, 
 are encoded in the collision terms ${\cal C}$, as explained in more detail in the next Section.

\subsection{Contributions to the Collision Term}
\label{sec:collision}

In this Section and throughout most of the remainder of this paper,
we consider the flavour relaxation rate
for left-handed SM leptons $\ell$. The methods employed and the calculations performed readily give rise to the production rate of
sterile right-handed neutrinos $N$, as it is explained in
Section~\ref{sec:rhproduction}. Numerical results of this
production rate together with a comparison to the earlier
results of Refs.~\cite{Anisimov:2010gy,Besak:2012qm}
are presented in Section~\ref{sec:rh:rate:results}.

\begin{figure}[t!]
\begin{center}
\epsfig{file=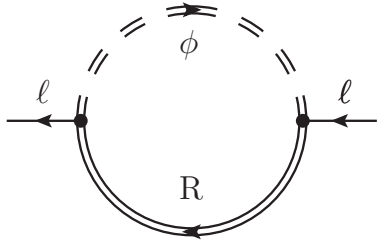,width=6cm}
\end{center}
\caption{
\label{fig:diagrams}
Diagram that represents the flavour-sensitive contribution
to the lepton self energy
${\Sigma\!\!\!/}^{(1)}_\ell$, that occurs at one-loop order
in the 2PI approach. Double lines indicate exact
(in the present approach one-loop resummed) propagators and the
external legs are understood to be amputated.
}
\end{figure}
The collision term for the lepton doublets,
\begin{align}
\label{collison:l}
	{\cal C}_\ell(p) & = {\rm i}{\Sigma\!\!\!/}_\ell^{>}(p) {\rm i}  S^<_\ell(p) - {\rm i}{\Sigma\!\!\!/}_\ell^{<}(p) {\rm i} S^>_\ell(p)\,,
\end{align}
can be calculated perturbatively by employing a loop expansion of the leptonic self energy ${\rm i}{\Sigma\!\!\!/}_\ell^{<,>}$. Furthermore,
the propagator ${\rm i} S_\ell^{<,>}$ contains self-energy insertions that
may be of importance and that
can also be expanded perturbatively.
In this paper, we restrict ourselves to
flavour-sensitive
contributions that are at least of order $h^2$ in the Yukawa couplings and that we
indicate by a superscript ${\rm fl}$, {\it i.e.} $\slashed\Sigma_\ell^{\rm fl}$
and ${\cal C}_\ell^{\rm fl}$. Due to the
non-trivial flavour structure, these couplings will give rise to flavour decoherence,
in contrast to the flavour-diagonal gauge couplings. The latter interactions are however
of importance, as they open up the phase space for flavour-decohering
processes (in combination with the coupling $h$)
that would be kinematically forbidden otherwise. Moreover,
gauge interactions maintain kinetic equilibrium for gauged particles,
which is what we assume throughout this work.

Up to two loops, there are contributions to the flavour-sensitive rates
from the one-loop self energy in Figure~\ref{fig:diagrams} and from the two-loop vertex-type diagrams in Figure~\ref{fig:fl_vertex}, which are obtained from the self-energy diagram by connecting two different propagators with a gauge boson. 

Note that in the 2PI approach,
no two-particle-reducible diagrams appear where a gauge boson connects to both
ends of the same propagator. Similarly, there is no diagrammatic contribution
from the insertion of a top-loop into the Higgs boson propagator.
While such diagrams with a self-energy insertion do not derive
explicitly from the 2PI loop expansion of the effective action, self-energy
insertions are of leading importance and are accounted for
implicitly, by using resummed propagators to evaluate the collision terms.
In fact, self-energy insertions in all propagators present in
Figure~\ref{fig:diagrams} contribute to the flavour decoherence at LO.
As we will see below, it is possible to expand the Higgs propagator up to the order of single loop insertions, whereas the lepton
propagators must be maintained in the fully resummed form due to the presence of
$t$-channel divergences from fermion exchange.

\begin{figure}[t!]
\begin{center}
\epsfig{file=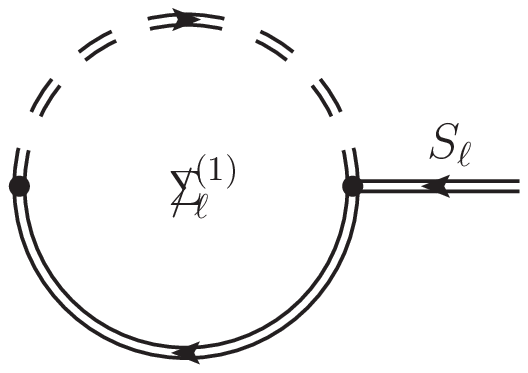,width=5cm}
\end{center}
\caption{
\label{fig:Cl:self}
The collision term ${\cal C}_\ell^{\rm self}$, Eq.~(\ref{C:self}).
}
\end{figure}

According to the loop expansion indicated in Figures~\ref{fig:diagrams}
and~\ref{fig:fl_vertex},
we decompose the collision term as
\begin{align}
	{\cal C}^{\rm fl}_\ell & = {\cal C}_\ell^{\rm self} + {\cal C}_\ell^{\rm vertex}.
\end{align}
The  contribution from the lepton self-energy diagram, Figure~\ref{fig:diagrams}, is given by 
\begin{align}
{\rm i}\slashed\Sigma_\ell^{{\rm (1)}ab}(p)
=&h^\dagger\int\frac{d^4k}{(2\pi)^4}\frac{d^4q}{(2\pi)^4}
(2\pi)^4\delta^4(p-k-q)
{\rm i}\Delta_\phi^{ab}(k){\rm i}S_{\rm R}^{ab}(q)h\,, \label{eqn:ell:selfen:1}
\end{align}
{\it i.e.}
\begin{align}
\label{C:self}
{\cal C}_\ell^{\rm self}={\rm i}{\Sigma\!\!\!/}_\ell^{(1)>}(p) {\rm i}S^<_\ell(p) - {\rm i}{\Sigma\!\!\!/}_\ell^{(1)<}(p) {\rm i}S^>_\ell(p)\,.
\end{align}
Within the 2PI approach, the propagators ${\rm i}S_{\ell,{\rm R}}$ and
${\rm i}\Delta_\phi$
are understood as the exact propagators. In the present work, we approximate these
by propagators that contain the resummed one-loop corrections that arise from
gauge interactions and from top-quark loops, {\it cf.} Section~\ref{sec:self}. Diagrammatically, the collison term~(\ref{C:self})
is given by Figure~\ref{fig:Cl:self}, where the double-line propagators
represent the resummed propagators.

When all propagators are on shell and massless, $p^2 = k^2 = q^2 = 0$, this self energy does not contribute to the collision term, so that no contribution of order $h^\dagger h$ arises. However once the medium effects are included by replacing the tree level propagators with resummed propagators, one obtains contributions that are parametrically of order $g^2 h^\dagger h$. The LO terms of this
type will be calculated in Section~\ref{sec:self}. 

The vertex-type diagrams give an order $g^2 h^\dagger h$ contribution to the collision term when evaluated using tree-level propagators. While in principle, it would be possible to evaluate also these diagrams with finite temperature resummed propagators, the difference between the two calculations is of higher order in the gauge coupling. For the purpose of the present work, it is therefore sufficient to calculate ${\cal C}_\ell^{\rm vertex}$ using tree level propagators, which will be performed in Section~\ref{sec:vertex}. 
%

In Ref.~\cite{Anisimov:2010gy} it is shown that collinearly enhanced
$1\leftrightarrow2$ processes appear as a third contribution that is parametrically
as important as the two contributions discussed above, and which requires a resummation of ladder diagrams with an arbitrary number of soft gauge bosons inserted between the Higgs and the lepton propagator in Figure~\ref{fig:diagrams}. These multiple scattering contributions
will be evaluated in Section~\ref{sec:onetwo}.

\subsection{Flavour Equilibration Rate}

The evolution equations for the left- (right-) handed lepton asymmetries
$q_\ell$ ($q_R$) are given by~\cite{flbg}
\begin{subequations}
\begin{align}
	\frac{d}{d t} q_{\ell}  &= -\left\{ W_\ell, q_\ell \right\} + 2 {\rm S}_\ell - \Gamma^{\rm fl}_\ell\,,
\\
\frac{d}{d t} q_{{\rm R}}  &= - \Gamma^{\rm fl}_{\rm R}\,, 
\end{align}
\end{subequations}
where in general, $q_{\ell, {\rm R}}$ and $W_\ell$, ${\rm S}_\ell$ and $\Gamma^{\rm fl}_{\ell,{\rm R}}$ are $3\times 3$ matrices in flavour space. The charge densities $q_{\ell,{\rm R}}$ are matrix-valued
generalisations of charge densities of left- and right-handed SM leptons, that can thereby
also account for flavour coherence. The lepton-number violating
washout term is denoted by $W_\ell$,
the source term for the $CP$-asymmetries by ${\rm S}_\ell$, and $\Gamma^{\rm fl}_{\ell,{\rm R}}$
describe the equilibration of lepton number between $\ell$ and ${\rm R}$ as well
as the decay of flavour coherence. In the standard scenarios of Leptogenesis,
there is no direct source for a charge asymmetry in right-handed leptons $R$
and no direct lepton-number violating coupling that woul induce a washout.

Here we are interested in the flavour equilibration rates $\Gamma^{\rm fl}_{\ell,{\rm R}}$ that depend on the Yukawa couplings $h_{ab}$. The rate $\Gamma^{\rm fl}_{\ell}$
can be obtained by integrating the relevant part of the collision term ${\cal C}_\ell$: 
\begin{align}
	-\Gamma^{\rm fl}_\ell & =\frac 12 \int \frac{d^4 p}{(2\pi)^4} {\rm tr}\,{\cal C}^{\rm fl}_\ell (p) + {\rm h.c.} =
	 \frac 12 \int \frac{d^4 p}{(2\pi)^4} 
{\rm tr}
	\left[
{\rm i}{\Sigma\!\!\!/}_{\ell}^{{\rm fl}>}(p) {\rm i}  S_{\ell}^{<}(p)
-{\rm i}{\Sigma\!\!\!/}_{\ell}^{{\rm fl}<}(p) {\rm i} S_{\ell}^{>}(p)
\right] +{\rm h.c.} \label{eqn:coll:ell}
\end{align}
where h.c. denotes hermitian conjugation in flavour space, and the superscript ${\rm fl}$ indicates that we only consider those contributions to the collision term that are at least second order in the charged lepton Yukawas $h_{ab}$. 
%
%
%
%
%
%
%
%

%
We assume that the particles $\ell$ and ${\rm R}$ are in kinetic equilibrium,
{\it i.e.} their distributions are of the Fermi-Dirac form with
generalised, matrix-valued chemical potentials $\mu_{\ell,{\rm R}\,ab}$.
Moreover, we assume that the chemical potentials are small,
$\mu_{\ell,{\rm R}\,ab}\ll T$. We can then expand the collision term to linear order in the chemical potentials using
\begin{align}
	f_{\ell ab}(k) &= \frac{1}{e^{\beta k^0}+1} \delta_{ab} + \delta f_{\ell ab}(k)\,, \\
	\delta f_{\ell ab}(k) &=  \beta \frac{e^{\beta k^0}}{(e^{\beta k^0}+1)^2} \mu_{\ell ab} \equiv 6 \beta^3 \frac{e^{\beta k^0}}{(e^{\beta k^0}+1)^2}  q_{\ell ab}\,.
\end{align}
Note that this also implies a decomposition of the Wigner functions  ${\rm i} S_\ell^{<,>} = {\rm i}  S_{\ell,{\rm eq}}^{<,>} + {\rm i} \delta S_\ell$, where the deviation from equilibrium is given by 
\begin{align}
{\rm i} \delta S_\ell = -2  S_\ell^{\cal A} \delta f_\ell(p) \,.
\end{align}
Accordingly,
the term linear in the right-handed SM
lepton asymmetry $q_{\rm R}$ is obtained by expanding ${\rm i} S_{\rm R}^{<,>}$, which enters $\Gamma^{\rm fl}_\ell$ through
Eqs.~(\ref{eqn:ell:selfen:1}) and~(\ref{eqn:coll:ell}). Now we expand
Eq.~(\ref{eqn:coll:ell}) in $\mu_{\ell,{\rm R}}/T$.
Due to the Kubo-Martin-Schwinger (KMS) relations,
${\rm i}  S_{\ell,{\rm R},{\rm eq}}^{>}(p)=-{\rm e}^{\beta p^0}{\rm i}   S_{\ell,{\rm R},{\rm eq}}^{<}(p)$
and ${\rm i} \slashed\Sigma_{\ell,{\rm eq}}^{{\rm fl}>}(p)=-{\rm e}^{\beta p^0}{\rm i} \slashed\Sigma_{\ell,{\rm eq}}^{{\rm fl}<}(p)$, where ${\rm i} \slashed\Sigma_{\ell,{\rm eq}}^{\rm fl}$ is the contribution from the equilibrium parts of the individual propagators to the self energy, the leading terms in this expansion are linear in
$\mu_{\ell,{\rm R}}$ ({\it i.e.} the equilibrium contributions vanish).
Therefore, the flavour-sensitive contribution
to the lepton collision term takes the general form
\begin{align}
\label{fl:rate}
-\frac{d q_\ell}{dt}  = 
\Gamma_\ell^{\rm fl} =
\frac{\gamma^{{\rm fl}\delta\ell}}{2} \left( h^\dagger h q_{\ell} + {\rm h.c.} \right)
+\frac{\gamma^{{\rm fl}\delta{\rm R}}}{2}  \left( h^\dagger q_{\rm R} h + {\rm h.c.} \right),
\end{align}
where we have defined the reduced scattering rates $\gamma^{{\rm fl} \delta \ell,R}$. 
Due to lepton number conservation of the interactions mediated by
$h$ in combination with flavour-blind gauge interactions, we
expect that
$\gamma^{{\rm fl}\delta\ell} = -\gamma^{{\rm fl}\delta{\rm R}} $,
what we explicitly verify in the calculations.
Note that
$q_{\ell ab} = \delta n^+_{\ell ab} - \delta n^-_{\ell ab}$ is the difference of the deviations of the lepton and anti-lepton number densities from thermal equilibrium. Fast gauge interactions ensure
that~\cite{flbg} $\delta n^+_{\ell ab}  = - \delta n^-_{\ell ab}$ such that $\Gamma^{\rm fl}_{\ell ab}$ can be obtained from the collision term according to
Eq.~(\ref{eqn:coll:ell}), otherwise the collision terms for particles and anti-particles would have to be evaluated separately, and  equations of motion for each $\delta n_{\ell ab}^+$ and $\delta n_{\ell ab}^-$ would have to be solved. 
%
%

The coefficients $\gamma^{{\rm fl}\delta\ell,\delta{\rm R}}$
have contributions from several diagrams,
\begin{align}
\label{gammas}
\gamma^{{\rm fl}\delta\ell,\delta{\rm R}}=
\gamma^{{\rm fl}{\rm(A)}\delta\ell,\delta{\rm R}}
+\gamma^{{\rm fl}{\rm(B)}\delta\ell,\delta{\rm R}}+\ldots
\,,
\end{align}
which we calculate in the following. Likewise, we
decompose the self energies $\slashed \Sigma_\ell$, the
collison terms ${\cal C}_\ell^{\rm fl}$ and the damping
rates $\Gamma_\ell^{\rm fl}$ into various diagrammatic components.

\subsection{Right-Handed Neutrino Production Rate}\label{sec:rhproduction}
The kinetic equation for the right-handed neutrino density $f_{N_i}(\mathbf{p})$ is given by
\begin{align}
	\frac{d}{d t} f_{N_i}(\mathbf p) & = D_i(\mathbf{p})\,, \label{eqn:eom:n}
\end{align}
where the decay (or inverse decay) rate $D_i$ is obtained by integrating the collision term over $p^0$~\cite{kblg}:
\begin{align}
  \label{eqn:coll:n}
	D_i(\mathbf p) & = \frac 14 \int \frac{dp^0}{2\pi} \sign(p^0) {\rm tr} {\cal C}_{Ni}(p) 
\\\notag
&=\frac 14 \int \frac{dp^0}{2\pi} \sign(p^0)
	{\rm tr}\left[
{\rm i}{\Sigma\!\!\!/}_{Nii}^{>}(p) {\rm i}S_{Ni}^{<}(p)
-{\rm i}{\Sigma\!\!\!/}_{Nii}^{<}(p) {\rm i}S_{Ni}^{>}(p)
\right].
\end{align}
Here, we neglect the effect of possible off-diagonal correlations in the
neutrino propagator. As long as the mass differences $M_{Ni}-M_{Nj}$ are much larger
than the relaxation rates $D$, this is a suitable approximation, because the
off-diagonal correlations oscillate rapidly and do not give a coherent contribution
to the diagonal evolution. In the case of a strong mass-degeneracy, the full evolution
of diagonal distributions as well as the oscillatory off-diagonal correlations must be
considered, as it is described in Ref.~\cite{Garbrecht:2011aw}. For simplicity,
we suppress in the following the flavour indices for the sterile
right-handed neutrinos.

For the right-handed neutrinos, the medium corrections to the propagator are proportional to $|Y|^2$ and therefore negligible in general, such that we can use the tree propagators~(\ref{spec:fun:fermi:tree}) with mass $M_N$ in the Wightman functions~(\ref{eqn:prop:fermion}). Inserting these into Eq.~(\ref{eqn:eom:n}), the $p^0$ integral can be performed analytically, and one obtains (suppressing the flavour indices)
\begin{align}
	\frac{d}{d t} f_{N}(\mathbf{p}) = \frac{1}{2 \omega(\mathbf{p})} {\rm tr}
	\left[ {p\!\!\!/} {\Sigma\!\!\!/}_{N}^{\cal A} (p)\right] 
	\left( \frac{1}{e^{\beta \omega(\mathbf{p})}+1} (1 - f_{N}(\mathbf{p})) - 
	\left(1 - \frac{1}{e^{\beta \omega(\mathbf{p})}+1}\right) f_{N}(\mathbf{p})  \right), \label{eqn:coll:n2}
\end{align}
where $p^0 = \omega(\mathbf p)=\sqrt{\mathbf p^2 +M_N^2}$ and where we have used that
$\ell$ and $\phi$ are in thermal equilibrium, such that the self energies satisfy the KMS condition ${\rm i} {\Sigma\!\!\!/}_N^{>}(p) = -e^{\beta p^0} {\rm i} {\Sigma \!\!\!/}_N^<(p)$. Furthermore,
the spectral part of the self energy is defined as
\begin{align}
	{\Sigma\!\!\!/}^{\cal A} = \frac{1}{2} \left( {\rm i} {\Sigma\!\!\!/}^> - {\rm i} {\Sigma\!\!\!/}^< \right).
\end{align}
Taking account of initial conditions for $f_{Ni}$ and of the expansion of the Universe,
solving Eq.~(\ref{eqn:coll:n2}) results in a non-equilibrium
distribution function $f_N(\mathbf p)$,
which can be computed numerically, but for which
no simple analytical form applies in general. Presenting an illustrative result
for the neutrino production rate therefore is to some extent a matter of
definition. For the ease of comparison, we follow Ref.~\cite{Anisimov:2010gy},
and choose the production rate for $f_N(\mathbf p)\equiv 0$. Approximately,
this is the relevant rate for the production of singlet neutrinos in the
weak washout scenario, provided it is assumed that initially, the
density of these particles vanishes. The differential production rate for the sum of the two
spin orientations is
\begin{align}
\label{totalrate}
\gamma^N=\frac{d\Gamma_N}{d^3p}=
\frac{2}{(2\pi)^3} \frac{d}{d t}
f_{N}(\mathbf p)
=
-\frac{1}{(2\pi)^3}\frac{1}{2p^0}
{\rm tr}[\slashed p{\rm i}\slashed\Sigma_N^<(p)]
\,,
\end{align}
where
\begin{align}
{\rm i}\slashed\Sigma^<_N(p)=
-2\frac{1}{{\rm e}^{\beta p^0}+1}\slashed\Sigma_N^{\cal A}(p)\,.
\end{align}
To see more explicitly how this rate is related to the flavour relaxation rate defined in the previous section, it is instructive to consider the leptonic collision term ${\cal C}^{\rm fl}_{\ell}(p)$. As before, we expand ${\rm i} S_\ell^{<,>} = {\rm i} S_{\ell,{\rm eq}}^{<,>} + {\rm i} \delta S_\ell$ and recall that ${\rm i} \delta S_\ell$ can be approximated
as being linear in $q_\ell$. To leading order in deviations from equilibrium, we can therefore 
assume that the self energies satisfy the KMS relation. This part of the collision term then simplifies to
\begin{align}
\frac{1}{2} \int \frac{dp^0}{2 \pi} {\rm tr} [{\cal C}^{\rm fl}_\ell(p)] = -\frac{1}{|p^0|}{\rm tr} [P_R {p\!\!\!/} {\Sigma\!\!\!/}^{{\rm fl}\cal A}_\ell(p) ] \delta f_\ell(p^0) + {\cal O}(\delta f_R)+{\rm h.c.}\,,
\end{align}
with $p^0 = \pm |\mathbf p|$. This has the same structure as the equation for the differential right-handed neutrino production rate (\ref{eqn:coll:n2}) except for the different statistical weight functions.
A relation between ${\rm tr}[\slashed p\slashed\Sigma_N^{\cal A}]$ and
${\rm tr}[\slashed p\slashed\Sigma_\ell^{{\rm fl}\cal A}]$ is obtained by isolating the dependence
on the coupling constants. In most cases, these are simple proportionalities,
whereas for $t$-channel fermion exchange, there is an additional logarithmic dependence
on the gauge couplings which can be isolated as well.
Therefore, once the differential (in $|\mathbf{p}|$) flavour relaxation rate is known, the $N$ production and decay rates can be obtained by simply performing the integral with the appropriate statistical weight and a rescaling of the couplings. 

\section{Self-Energy Type Contributions}
\label{sec:self}

\subsection{One-Loop Self Energies}

The resummed form of the spectral function of a
massless chiral fermion is~\cite{flbg,Prokopec:2003pj,Garbrecht:2008cb}
\begin{align}
\label{S:spectral:resummed}
 S^{\cal A}(k)
=P_{X}\frac{
2 \left(k\!\!\!/-{\Sigma\!\!\!\!/}^H\right)
\Sigma^{\cal A}\cdot(k-\Sigma^H)
-{\Sigma\!\!\!\!/}_\ell^{\cal A}\left(k\!\!\!/-{\Sigma\!\!\!\!/}^H\right)^2
+{{{{\Sigma\!\!\!\!/}}^{{\cal A}\,3}}} 
}
{
\left[
\left(k\!\!\!/-{\Sigma\!\!\!\!/}^H\right)^2
-{{\Sigma\!\!\!\!/}^{{\cal A}\,2}}
\right]^2
+4\left[
\Sigma^{\cal A}\cdot(k-\Sigma^H)
\right]^2
}
\,,
\end{align}
and of a massless scalar
\begin{align}
\label{specfun:Higgs}
\Delta_\phi^{\cal A}(k)
=\frac{\Pi^{\cal A}_\phi(k)}{k^4+[{\Pi^{\cal A}_\phi(k)}]^2}\,.
\end{align}

We express the spin-$\frac12$ fermionic self energies as
$\slashed\Sigma=\gamma^\mu\Sigma_\mu$. In the approximation
of massless particles in the loop, the spectral self energy
for leptons $\ell$ or ${\rm R}$
is given by
\begin{subequations}
\label{Sig:A}
\begin{align}
\Sigma^{{\cal A}0}_{\ell,{\rm R}}(k)
&=\frac{G T^2}{16\pi|\mathbf k|}
I_1\left({\frac{k^0}{T},\frac{|\mathbf k|}{T}}\right)\,,\\
\Sigma^{{\cal A}i}_{\ell,{\rm R}}(k)
&=\frac{G T^2}{16\pi|\mathbf k|}
\left[
\frac{k^0}{|\mathbf k|}
I_1\left({\frac{k^0}{T},\frac{|\mathbf k|}{T}}\right)
-\frac{(k^0)^2-\mathbf k^2}{2|\mathbf k|T}
I_0\left({\frac{k^0}{T},\frac{|\mathbf k|}{T}}\right)
\right]\frac{k^i}{|\mathbf k|}\,,
\end{align}
\end{subequations}
where
\begin{subequations}
\begin{align}
I_0(y^0,y)=&-\vartheta(y^2-(y^0)^2)y^0-y
+\log
\left|
\frac{1+{\rm e}^{\frac12(y^0+y)}}{1+{\rm e}^{\frac12(y^0-y)}}
\right|
+\log
\left|
\frac{1-{\rm e}^{\frac12(y^0+y)}}{1-{\rm e}^{\frac12(y^0-y)}}
\right|\,,
\\
I_1(y^0,y)=&-\vartheta(y^2-(y^0)^2)\frac{(y^0)^2-\pi^2}{2}
\\\notag
+&{\rm Re}\left[
x(\log(1+{\rm e}^x)-\log(1-{\rm e}^{x-y^0}))
+{\rm Li}_2(-{\rm e}^{x})-{\rm Li}_2({\rm e}^{x-y^0})
\right]^{x=\frac12(y^0+y)}_{x=\frac12(y^0-y)}
\,.
\end{align}
\end{subequations}
For the case of the Standard Model lepton doublet $\ell$,
$G=\frac12(3 g_2^2+g_1^2)$, whereas for right-handed leptons ${\rm R}$,
$G=2 g_1^2$.
Notice that these values for $G$ include a factor of two
that accounts for the polarisation states of the gauge bosons,
as the spectral self energies~(\ref{Sig:A}) are defined here
for a single bosonic and a fermionic (with both spin states) degree
of freedom in the loop.
Note as well
that we distinguish between the lepton self energy~(\ref{Sig:A}),
that is flavour-diagonal and of order $g^2$, and the flavour-sensitive self energy
${\rm i}\slashed\Sigma_\ell^{\rm fl}$, for which the
LO terms are $\sim h^2 g^2$ and $\sim h^2 g^2 \log g^2$.

The spectral self energy for the Higgs field is given by
\begin{subequations}
\label{Pi:spectral:phi}
\begin{align}
\Pi^{\cal A}_\phi
=&\frac{\frac32g_2^2+\frac12g_1^2}{16\pi}\frac{k^2}{|\mathbf k|}
\left(
|\mathbf k|
-\frac{2}{\beta}\log
\frac{1-{\rm e}^{\beta\frac{k^0+|\mathbf k|}{2}}}
{1-{\rm e}^{\beta\frac{k^0-|\mathbf k|}{2}}}
\right)
\\
\nonumber
-&\frac{3h_t^2}{16\pi}\frac{k^2}{|\mathbf k|}
\left(
|\mathbf k|
-\frac{2}{\beta}\log
\frac{1+{\rm e}^{\beta\frac{k^0+|\mathbf k|}{2}}}
{1+{\rm e}^{\beta\frac{k^0-|\mathbf k|}{2}}}
\right)
\;\;\textnormal{for}\;\;k^2\geq0\,,
\end{align}
\begin{align}
\Pi^{\cal A}_\phi
=&\frac{\frac32g_2^2+\frac12g_1^2}{16\pi}\frac{k^2}{|\mathbf k|}
\left(
2k^0
-\frac{2}{\beta}\log
\frac{1-{\rm e}^{\beta\frac{|\mathbf k|+k^0}{2}}}
{1-{\rm e}^{\beta\frac{|\mathbf k|-k^0}{2}}}
\right)
\\
\nonumber
-&\frac{3h_t^2}{16\pi}\frac{k^2}{|\mathbf k|}
\left(
2k^0
-\frac{2}{\beta}\log
\frac{1+{\rm e}^{\beta\frac{|\mathbf k|+k^0}{2}}}
{1+{\rm e}^{\beta\frac{|\mathbf k|-k^0}{2}}}
\right)
\;\;
\textnormal{for}\;\;k^2<0\,.
\end{align}
\end{subequations}

When we substitute tree-level propagators for all three fields $\ell$, ${\rm R}$
and $\phi$ that appear in the expression for ${\cal C}_\ell^{\rm self}$
defined by Eqs.~(\ref{eqn:ell:selfen:1},\ref{C:self}), we obtain a
vanishing result, because the $1\leftrightarrow 2$ process is kinematically
forbidden (when neglecting the tree-level masses) without including
the finite-temperature corrections. At leading order in the gauge and top-quark
Yukawa couplings, the medium corrections add linearly such that
the one-loop (in the 2PI sense) self energy~(\ref{eqn:ell:selfen:1})
for the lepton doublet can be approximated as
\begin{align}
{\rm i}\slashed\Sigma_\ell^{(1)}(p)
\approx
{\rm i}\slashed\Sigma_\ell^{(\rm R)}(p)
+{\rm i}\slashed\Sigma_\ell^{(\phi)}(p)
\qquad\textnormal{for}\;p^2=0
\,,
\end{align}
where
\begin{subequations}
\begin{align}
{\rm i}\slashed\Sigma_\ell^{({\rm R})ab}(p)
=&h^\dagger\int\frac{d^4k}{(2\pi)^4}\frac{d^4q}{(2\pi)^4}
(2\pi)^4\delta^4(p-k-q)
{\rm i}\Delta_\phi^{(0)ab}(k){\rm i} S_{\rm R}^{ab}(q)h\,,
\\
{\rm i}\slashed\Sigma_\ell^{(\phi)ab}(p)
=&h^\dagger\int\frac{d^4k}{(2\pi)^4}\frac{d^4q}{(2\pi)^4}
(2\pi)^4\delta^4(p-k-q)
{\rm i}\Delta_\phi^{ab}(k){\rm i} S_{\rm R}^{(0)ab}(q)h
\,,
\\
{\rm i}\slashed\Sigma_\ell^{({\rm os})ab}(p)
=&h^\dagger\int\frac{d^4k}{(2\pi)^4}\frac{d^4q}{(2\pi)^4}
(2\pi)^4\delta^4(p-k-q)
{\rm i}\Delta_\phi^{(0)ab}(k){\rm i} S_{\rm R}^{(0)ab}(q)h\,.
\end{align}
\end{subequations}
Here, we have also defined the contribution from on-shell
Higgs bosons and right-handed SM leptons
${\rm i}\slashed\Sigma_\ell^{({\rm os})ab}(p)$, where
${\rm i}\slashed\Sigma_\ell^{({\rm os})<,>}(p)=0$ for $p^2=0$
for the kinematic reasons mentioned above. However, since
the resummed propagator ${\rm i} S_\ell^{<,>}(p)$ is
non-vanishing for $p^2\not=0$, there occurs a contribution
involving ${\rm i}\slashed\Sigma_\ell^{({\rm os})ab}(p)$ that
is kinematically allowed due to gauge bosons that may radiate
from $\ell$.
\begin{figure}[t!]
\begin{center}
\epsfig{file=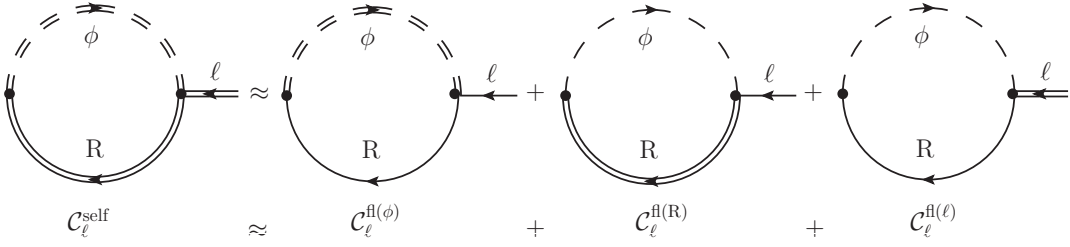,width=15cm}
\end{center}
\caption{
\label{fig:Cell:decomp}
Diagrammatic representation of the LO expansion of the one-loop
2PI collision term, Eq.~(\ref{C:ell:self:decompose}).
}
\end{figure}
In summary, we can decompose the self-energy contributions to
the flavour-decohering collision term as
\begin{align}
\label{C:ell:self:decompose}
{\cal C}_\ell^{\rm self}\approx{\cal C}_\ell^{{\rm fl}(\phi)}+{\cal C}_\ell^{{\rm fl}({\rm R})}+{\cal C}_\ell^{{\rm fl}(\ell)}\,,
\end{align}
where
\begin{align}
\label{C:ell:self:decompose:2}
{\cal C}_\ell^{{\rm fl}(\phi,{\rm R})}=&{\rm tr}
\left[
{\rm i}\slashed\Sigma_\ell^{(\phi,{\rm R})>}(p){\rm i}S_\ell^{(0)<}(p)
-{\rm i}\slashed\Sigma_\ell^{(\phi,{\rm R})<}(p){\rm i}S_\ell^{(0)>}(p)
\right]\,,
\\
{\cal C}_\ell^{{\rm fl}(\ell)}=&{\rm tr}\left[
{\rm i}\slashed\Sigma_\ell^{({\rm os})>}(p){\rm i}S_\ell^<(p)
-{\rm i}\slashed\Sigma_\ell^{({\rm os})<}(p){\rm i}S_\ell^>(p)
\right]\,.
\end{align}
A diagrammatic representation of the
decomposition~(\ref{C:ell:self:decompose}) is given by
Figure~\ref{fig:Cell:decomp}.

\subsection{Scatterings via the Higgs Boson}

Substituting the tree-level spectral function~(\ref{spec:fun:fermi:tree}) for $\ell$ and
${\rm R}$ and the one-loop resummed spectral function for
$\phi$, the collision term becomes
\begin{align}
\label{C:wv:phi}
{\cal C}_\ell^{{\rm fl}(\phi)}
=&\int\frac{d^4k}{(2\pi)^4}\frac{d^4q}{(2\pi)^4}
(2\pi)^4\delta^4(p+k-q)
\frac{2\Pi_\phi^{\cal A}(k)}{k^4+[\Pi^{\cal A}_\phi(k)]^2}
2p\cdot q
\\\notag
\times&
2\pi\delta(p^2)2\pi\delta(q^2){\rm sign}(p^0){\rm sign}(q^0)
\\\notag
\times&
h^\dagger
\left[
(1+f_\phi(k^0)) f_{\rm R}(q^0) h (1-f_\ell(p^0))
-f_\phi(k^0)(1-f_{\rm R}(q^0)) h f_\ell(p^0)
\right]\,.
\end{align}
It is now useful to notice that $\Pi^{\cal A}(k)$ is first order
in $k^2$ for $k^2\to 0$
[{\it cf.} Eq.~(\ref{Pi:spectral:phi})], as well as $2 p\cdot q=k^2$
(due to the relations imposed by the $\delta$-functions).
To leading order in the couplings $g^2_{1,2}$ and $h_t^2$,
we may therefore replace
\begin{align}
\label{repl:noresum}
\frac{2\Pi_\phi^{\cal A}}{k^4+[\Pi^{\cal A}_\phi]^2}
\approx
\frac{2\Pi_\phi^{\cal A}}{k^4}\,,
\end{align}
such that in this approximation, ${\cal C}_\ell^{{\rm fl}(\phi)}$
is proportional to $g^2_{1,2}$ and $h_t^2$. Alternatively,
one can derive Eq.~(\ref{C:wv:phi})
with the replacement~(\ref{repl:noresum}) from
a CTP two-loop (two particle reducible)
self energy in terms of tree-level propagators.

We now assume that $\phi$, $\ell$, ${\rm R}$ are in kinetic equilibrium,
and that the chemical potentials of $\ell$ and ${\rm R}$ are small
compared to the temperature, such that we may relate
\begin{align}
\delta f_{\ell,{\rm R}}(p^0)=
6\beta^3\frac{{\rm e}^{\beta p^0}}{({\rm e}^{\beta p^0}+1)^2}
q_{\ell,{\rm R}}
\,,
\end{align}
where
$\delta f_{\ell,{\rm R}}(p^0)=f_{\ell,{\rm R}}(p^0)-f_F^{\rm eq}(p^0)$.
Note that $\delta f_{\ell,{\rm R}}(p^0)$ and $q_{\ell,{\rm R}}$
are understood as matrices in flavour space and $f_F^{\rm eq}(p^0)$,
the Fermi-Dirac equilibrium distribution for vanishing chemical potential,
is therefore implied to be proportional to the unit matrix.

With the definitions~(\ref{eqn:coll:ell},\ref{fl:rate},\ref{gammas}), it follows
\begin{align}
\gamma^{{\rm fl}(\phi)\delta\ell} 
=&-\gamma^{{\rm fl}(\phi)\delta{\rm R}} \notag \\
=&-\int\frac{d^4 k}{(2\pi)^4}
\int\frac{d^4 p}{(2\pi)^4}
\int\frac{d^4 q}{(2\pi)^4}
(2\pi)^4\delta^4(p+k-q)\frac{2\Pi^{\cal A}(k)}{k^2}
2\pi\delta(p^2)2\pi\delta(q^2)
\\\notag
\times&
{\rm sign}(p^0){\rm sign}(q^0)
\frac{
{\rm e}^{\beta k^0}+{\rm e}^{\beta q^0}
}
{
({\rm e}^{\beta k^0}-1)({\rm e}^{\beta q^0}+1)
}
6\beta^3\frac{{\rm e}^{\beta p^0}}{({\rm e}^{\beta p^0}+1)^2}
\\\notag
=&7.71\times10^{-4}\times \left(\frac32 g_2^2 +\frac12 g_1^2\right)T
+1.32\times10^{-3}\times h_t^2 T
\,,
\end{align}
where the last expression is the result of numerical evaluation of
the integrals.

\subsection{Scatterings via Fermions}\label{sec:fermions}

Now consider the terms that arise from substituting the tree-level
spectral function~(\ref{specfun:scalar:tree}) for the scalar propagator
and~(\ref{spec:fun:fermi:tree}) for one of the fermions $\ell$ or ${\rm R}$,
while using the resummed spectral function~(\ref{S:spectral:resummed}) for the other fermion.
For definiteness, we calculate the collison term for scatterings
via ${\rm R}$ first
[{\it i.e.} with ${\rm i}S_{\rm R}^{\cal A}$ as in
Eq.~(\ref{S:spectral:resummed})]:
\begin{align}
\label{C:wv:R}
{\cal C}_\ell^{{\rm fl}({\rm R})}(p)
=&\int\frac{d^4k}{(2\pi)^4}\frac{d^4q}{(2\pi)^4}
(2\pi)^4\delta^4(p-k+q)
2 {\rm tr}\left[S_{\rm R}^{\cal A}(k)\slashed p \right]
\\\notag
\times&
2\pi\delta(p^2)2\pi\delta(q^2){\rm sign}(p^0){\rm sign}(q^0)
\\\notag
\times&
\left[
h^\dagger f_{\rm R}(k^0) h (1-f_\ell(p^0))(1+f_\phi(q^0))
-h^\dagger (1-f_{\rm R}(q^0)) h f_\ell(p^0)f_\phi(q^0)
\right]
\,.
\end{align}
The contribution from scatterings via $\ell$ can
directly be inferred from the evaluation of this term.
Expanding in the deviations of $f_{\ell,{\rm R}}$ from
chemical equilibrium, we write
\begin{align}
{\cal F}(k^0,p^0,q^0)
=&
\left[
h^\dagger f_{\rm R}(k^0)h(1-f_\ell(p^0))(1+f_\phi(q^0))
-h^\dagger(1-f_{\rm R}(q^0)) h f_\ell(p^0)f_\phi(q^0)
\right]
\\\notag
=&-h^\dagger h\delta f_\ell(p^0)\left[\frac{1}{{\rm e}^{\beta k^0}+1}+\frac{1}{{\rm e}^{\beta q^0}-1}\right]
\\\notag
+&h^\dagger \delta f_{\rm R} (k^0) h
\left[1-\frac{1}{{\rm e}^{\beta p^0}+1}+\frac{1}{{\rm e}^{\beta q^0}-1}\right]\,,
\end{align}
what leads to the decomposition
${\cal C}_\ell^{{\rm fl}({\rm R})}
={\cal C}_\ell^{{\rm fl}({\rm R})\delta\ell}
+{\cal C}_\ell^{{\rm fl}({\rm R})\delta{\rm R}}$ and accordingly
for $\Gamma^{\rm fl}$ and $\gamma^{\rm fl}$.

The collision term~(\ref{C:wv:R}) can straightforwardly be evaluated 
numerically. After making use of the $\delta$-functions,
homogeneity and isotropy, two numerical integrations are
left. However, it is useful and instructive to isolate
the dependence on the coupling $G$.

It is known that the phase space integrals over the scattering
matrix elements exhibit a logarithmic divergence for
{\it zero}-momentum exchange of a lepton ${\rm R}$ in the
$t$ channel. The scattering approximation is recovered
in the present approach when omitting the self energies
in the denominator of the spectral function, Eq.~(\ref{S:spectral:resummed}), which
would lead to a logarithmic divergence in the
integral~(\ref{C:wv:R}) for $k^0=|\mathbf k|=0$ and
$k^2<0$. A simplification corresponding to
the replacement~(\ref{repl:noresum}) is therefore not suitable
for the present integral.
A finite part can however be extracted when subtracting
those terms from the spectral self energy
$\slashed\Sigma^{\cal A}_{\rm R}$ that are not vanishing in the
limit $k^0,|\mathbf k|\to0$. These are precisely the contributions that
are of the form of the hard thermal loop (HTL) approximation and that we
indicate by a tilde. We therefore define
\begin{align}
\bar I_1(y^0,y)&=I_1(y^0,y)-\tilde I_1(y^0,y)\,,
\\
\tilde I_1(y^0,y)&=
\vartheta(y^2-(y^0)^2)\frac{\pi^2}{2}\,.
\end{align}
The barred quantities $\bar S^{\cal A}$ and
$\bar \Sigma$
are defined through the replacement $I_1\to \bar I_1$, and
accordingly
$\Sigma^{\cal A}=\bar \Sigma^{\cal A}+\tilde\Sigma^{{\cal A}}$,
${\cal C}_\ell^{{\rm fl}{\rm (R)}}
=\bar {\cal C}_\ell^{{\rm fl}{\rm (R)}}+\tilde {\cal C}_\ell^{{\rm fl}{\rm (R)}}$. Note that $I_0$ vanishes in the HTL approximation.

For the purpose of calculating $\bar {\cal C}_\ell^{{\rm fl}{\rm (R)}}$ to
leading order in $G$, it is sufficient to approximate
\begin{align}
\label{eq:prop:nonhtl}
\bar S_{\rm R}^{\cal A}(k)
=P_{\rm R}\frac{
2 k\!\!\!/
\bar\Sigma_{\rm R}^{\cal A}\cdot k
-\bar{\Sigma\!\!\!\!/}_{\rm R}^{\cal A} k^2
}
{
k^4
}
P_{\rm L}
\,.
\end{align}
The result for $\bar {\cal C}_\ell^{{\rm fl}{\rm (R)}}$
is therefore
manifestly proportional to $G$.

It remains to calculate the part that originates
from the term of the HTL form. Therefore, we need to find
\begin{align}
\tilde{\cal C}_\ell^{{\rm fl}{\rm (R)}}(p)=&
2\pi\delta(p^2){\rm sign}(p^0)\int\frac{d^4k}{(2\pi)^4}\frac{d^4q}{(2\pi)^4}
(2\pi)^4\delta(p-k+q)
2\pi\delta(q^2){\rm sign}(q^0)
\\\notag
\times&2{\rm tr}
\left[
\slashed p
P_{\rm R}
\frac
{
2(\slashed k-\slashed\Sigma_{\rm R}^H)
\tilde\Sigma_{\rm R}^{\cal A}\cdot(k-\Sigma_{\rm R}^H)
-\tilde{\slashed\Sigma}_{\rm R}^{\cal A}
\left(\slashed k-\slashed\Sigma_{\rm R}^H\right)^2
+{\tilde{\slashed\Sigma}_{\rm R}^{\cal A}{}}^3
}
{
\left[
\left(\slashed k-\slashed\Sigma_{\rm R}^H\right)^2
-{{{\tilde{\slashed\Sigma}}_{\rm R}^{\cal A}}{}}^2
\right]^2
+4
\left[
\tilde \Sigma^{\cal A}\cdot(k-\Sigma^H)
\right]^2
}
\right]
\times{\cal F}(k^0,p^0,q^0)
\,.
\end{align}

In order to extract the dependence on $G$, it is useful to
split the integral in a region where
$|\mathbf k|\geq k_*\gg \sqrt G T$, where the denominator simplifies
(because the self energies may be neglected there far from the
single-particle poles)
and a region where $|\mathbf k|\leq k_*\ll|\mathbf p|$, where
the angular integration simplifies,
$\tilde{\cal C}_\ell^{{\rm fl}{\rm (R)}}(p)=
\tilde{\cal C}_{\ell,<k_*}^{{\rm fl}{\rm (R)}}(p)
+\tilde{\cal C}_{\ell,>k_*}^{{\rm fl}{\rm (R)}}(p)
$. Notice that as a consequence of this split, the result should
only be valid when $|\mathbf p|\gg\sqrt{GT}$. Due to the phase space
suppression, this does however not spoil the LO calculation of
the flavour relaxation rate.

Dropping the terms $\propto G$ in the denominator, we evaluate
\begin{align}
\tilde{\cal C}^{{\rm fl}{\rm (R)}}_{\ell,>k_*}(p)
&\underset{p^0=|\mathbf p|}{=}\frac{GT^2}{2^7 \pi}2\pi\delta(p^2)
\int\limits_{-1}^1d\cos\vartheta\int\limits_{k_*}^\infty \frac{d|\mathbf k|}{|\mathbf k|}
\frac{k^0-2|\mathbf p|}{|\mathbf p|-k^0}
{\cal F}(k^0,p^0,k^0-p^0)\,,
\end{align}
where
\begin{align}
k^0=|\mathbf p|-\sqrt{\mathbf p^2+\mathbf k^2-2|\mathbf p||\mathbf k|\cos\vartheta}\,.
\end{align}
(Note that the HTL contributions are only present for
$k^2<0$.)
The collision term for $p^0<0$ may be obtained when noting that it
is even in $p^0$, provided the particle distributions are in
kinetic equilibrium.

The logarithmic dependence can be isolated through integration by parts. We simplify the
angular integration and additional terms using  $k_*\ll|\mathbf p|$,
such that $k^0\approx|\mathbf k|\cos\vartheta$, and obtain
\begin{subequations}
\begin{align}
\tilde{\cal C}^{{\rm fl}{\rm (R)}}_{\ell,>k_*}
&=\tilde{\cal C}^{{\rm fl}{\rm (R)}}_{\ell,>k_*{\rm LOG}}
+\tilde{\cal C}^{{\rm fl}{\rm (R)}}_{\ell,>k_*{\rm FIN}}\,,
\\
\tilde{\cal C}^{{\rm fl}{\rm (R)}}_{\ell,>k_*{\rm LOG}}
&=\frac{G T^2}{2^5 \pi}2\pi\delta(p^2)\log(\beta k_*){\cal F}(0,p^0,-p^0)
\,,\\
\tilde{\cal C}^{{\rm fl}{\rm (R)}}_{\ell,>k_*{\rm FIN}}
&=\frac{G T^2}{2^7 \pi}2\pi\delta(p^2)
\int\limits_{-1}^1d\cos\vartheta\int\limits_{k_*}^\infty
d|\mathbf k|\log(\beta |\mathbf k|)\frac{\partial}{\partial(\beta|\mathbf k|)}
\frac{k^0-2|\mathbf p|}{k^0-|\mathbf p|}
{\cal F}(k^0,p^0,k^0-p^0)
\,,
\end{align}
\end{subequations}
which we have separated into a contribution
$\tilde{\cal C}^{{\rm fl}{\rm (R)}}_{\ell,>k_*{\rm LOG}}$
that depends logarithmically on $k_*$
and an integral
$\tilde{\cal C}^{{\rm fl}{\rm (R)}}_{\ell,>k_*{\rm FIN}}$
that depends only linearly on $k_*$ ({\it i.e.}
that is finite for $k_*\to 0$), such that we can
take the lower bound of the integration to zero when $k_*\ll|\mathbf p|$.

In order to calculate ${\cal I}_{<p_*}$, it is necessary to take account of the
screening that is induced by the self energies. In addition
to the spectral self energy, also the hermitian part is of
importance. Because $p_*\ll T$, it is sufficient to consider the
HTL approximations
\begin{subequations}
\begin{align}
\Sigma^{H0}(k)&=\frac{G T^2}{32|\mathbf k|}\log\left|\frac{k^0+|\mathbf k|}{k^0-|\mathbf k|}\right|\,,\\
\Sigma^{Hi}(k)&=\frac{G T^2 k^0 k^i}{32|\mathbf k|^3}\log\left|\frac{k^0+|\mathbf k|}{k^0-|\mathbf k|}\right|-\frac{G T^2 k^i}{16 \mathbf k^2}\,.
\end{align}
\end{subequations}

We approximate
$|\mathbf p-\mathbf k|=|\mathbf p|-\hat{\mathbf p}\cdot \mathbf k$, and moreover, we
evaluate the statistical functions for $k^0=|\mathbf k|\approx 0$, as it is appropriate
for $k_*\ll |\mathbf p|$. Numerically, we can then obtain the value of
\begin{align}
\tilde{\cal C}^{\prime{\rm fl}{\rm(R)}}_{\ell,<k_*^\prime}
=
-2\pi\delta(p^2)\frac{{\cal F}(0,p^0,-p^0)}{(2\pi)^2|\mathbf p|}
\int\limits_{-1}^1d\cos\vartheta
\int\limits_0^{k^\prime_*}
\mathbf k^2d|\mathbf k|{\rm tr}
\left[\slashed p \tilde S_{\rm R}^{\prime {\cal A}}(k)\right]
\,.
\end{align}
The prime on $\tilde S_{\rm R}^{\prime {\cal A}}(k)$ indicates that
we evaluate this expression by replacing $G\to G^\prime$, and the
tilde indicates, that we use the HTL approximation.
Above expression then corresponds to
the infrared contribution to the scattering rates that arises from a UV cutoff $k_*^\prime$
and a squared coupling $G^\prime$. From the dependence of
the HTL self energies on $G$ and on the four-momentum, we observe
that a simultaneous rescaling of the squared
coupling by a factor of $G/G^\prime$
and of the momentum by $\sqrt{G/G^\prime}$ rescales
the value of the integrand (including the integration
measure) by an overall factor
of $G/G^\prime$. Therefore, $(G/G^\prime) \tilde{\cal C}^{\prime{\rm fl}({\rm R})}_{<k_*^\prime}$
also describes scatterings for a squared
coupling $G$ and a UV cutoff $\sqrt{G/G^\prime} k_*^\prime$.
In order to obtain the contribution for a cutoff $k_*$, we
must add the integral
\begin{align}
\tilde{\cal C}^{{\rm fl}{\rm(R)}}_{\ell[\sqrt{G/G^\prime} k_*^\prime,k_*]}&=
-2\pi\delta(p^2)\frac{{\cal F}(0,p^0,-p^0)}{(2\pi)^2|\mathbf p|}
\int\limits_{-1}^1d\cos\vartheta
\int\limits_{\sqrt{G/G^\prime} k_*^\prime}^{k_*}
\mathbf k^2d|\mathbf k|{\rm tr}
\left[\slashed p \tilde S_{\rm R}^{{\cal A}}(k)\right]
\\\notag
&=
-2\pi \delta(p^2) \frac{G T^2}{2^6 \pi}{\cal F}(0,p^0,-p^0)
\log
\left(
\frac{G^\prime k_*^2}{G {k_*^\prime}^2}
\right)
\,.
\end{align}

In summary, the wave-function contribution to the scattering rate
can be decomposed as
\begin{align}
\label{C:ell:semi-analytical}
{\cal C}^{{\rm fl}{\rm (R)}}_\ell
=\bar{\cal C}^{{\rm fl}{\rm (R)}}_\ell
+\tilde{\cal C}^{{\rm fl}{\rm (R)}}_{\ell,>k_*{\rm LOG}}
+\tilde{\cal C}^{{\rm fl}{\rm (R)}}_{\ell,>k_*{\rm FIN}}
+\tilde{\cal C}^{{\rm fl}{\rm (R)}}_{\ell[\sqrt{G/G^\prime}k_*^\prime,k_*]}
+\frac{G}{G^\prime}\tilde{\cal C}^{\prime{\rm fl}{\rm (R)}}_{\ell,<k^\prime_*}
\,.
\end{align}
The terms $\bar{\cal C}^{{\rm fl}{\rm (R)}}_\ell$,
$\tilde{\cal C}^{{\rm fl}{\rm (R)}}_{\ell,>k_*{\rm FIN}}$
and
$\frac{G}{G^\prime}\tilde{\cal C}^{\prime{\rm fl}{\rm (R)}}_{\ell,<k^\prime_*}$ should be evaluated
numerically and are proportional to $G$.
The logarithmic dependence on $G$, that results from the screening
of scattering processes with small momentum exchange is isolated
in
\begin{align}
\tilde{\cal C}^{{\rm fl}{\rm (R)}}_{\ell,>k_*{\rm LOG}}
+\tilde{\cal C}^{{\rm fl}{\rm (R)}}_{\ell[\sqrt{G/G^\prime}k_*^\prime,k_*]}
=2\pi\delta(p^2)\frac{G T^2}{2^6 \pi}
\log\left(\frac{G}{G^\prime}\beta^2{k_*^\prime}^2\right)
{\cal F}(0,p^0,-p^0)
\,.
\end{align}
We emphasise that the final result
${\cal C}^{{\rm fl}{\rm (R)}}_\ell$
is by construction independent
on the choice of $k_*$, $k_*^\prime$ and $G^\prime$,
as the dependence
of above expression on these parameters is compensated by
$\frac{G}{G^\prime}\tilde{\cal C}^{\prime{\rm fl}{\rm (R)}}_{\ell,<k^\prime_*}$.
Recall that this approximate
cancellation of the dependence on $k_*$, $k_*^\prime$ and $G^\prime$ is a consequence
of the approximate
behaviour of the integrand when $G\sqrt{T}\ll|\mathbf k|\ll|\mathbf p|$.
Below, we verify this numerically in order to test the accuracy of the
approximations.

\begin{figure}[t!]
\begin{center}
\epsfig{width=10cm, file=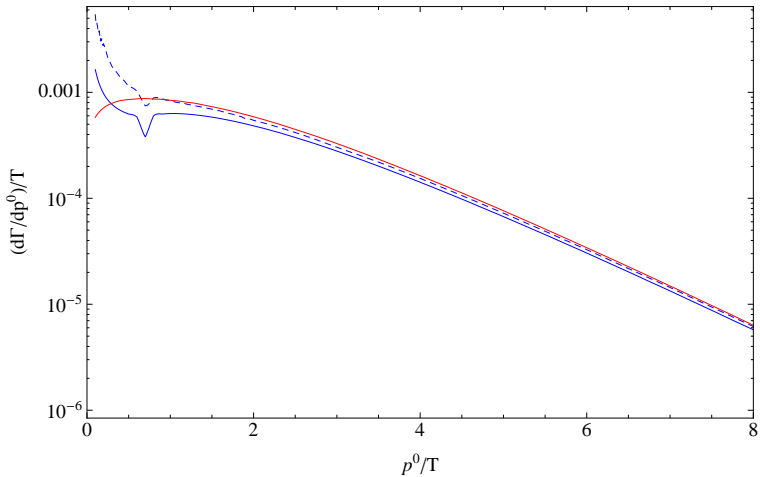}
\end{center}
\caption{
\label{fig:kSigma}
${\cal C}^{{\rm fl}{\rm (R)}}_\ell (p)p^0/(\delta q_\ell) 2\pi \delta(p^2)$ for $\delta q_{\rm R}=0$ over
$p^0=|\mathbf p|$ ($p^2=0$) for
$G=0.3$. Blue: semi-analytical, Eq~(\ref{C:ell:semi-analytical}). Red: full integral. Blue, dashed: semi-analytical without the approximation~(\ref{eq:prop:nonhtl}).
}
\end{figure}

Scatterings may as well proceed via the exchange of a doublet
lepton $\ell$. This contribution to the collision term is
${\cal C}_\ell^{{\rm fl}({\ell})}(p)$, Eq.~(\ref{C:ell:self:decompose:2}).
With the integration over $d^4 p$, the relevant
integrals are identical to those for the exchange of ${\rm R}$. From
the result for the scattering via ${\rm R}$, we can therefore
directly infer the contribution from exchanges of $\ell$.

Substituting the collision term ${\cal C}^{{\rm fl}{\rm (R)}}_\ell$
into the expression for the relaxation rate~(\ref{fl:rate})
and using the definition~(\ref{eqn:coll:ell}), we find
\begin{align}
\label{gamma:t-fermi:semi-an}
\gamma^{{\rm fl}({\rm R},\ell)\delta\ell}
=-\gamma^{{\rm fl}({\rm R},\ell)\delta{\rm R}}
=4.40\times 10^{-3} \times G T - 9.33\times 10^{-4} \times G T \log G\,,
\end{align}
where $G=2 g_1^2$ for ${\rm R}$ exchange (superscript $({\rm R})$)
and $G=\frac32 g_2^2 + \frac12 g_1^2$
for $\ell$ exchange (superscript $(\ell)$).
Notice that there is an analytical expression for the numerical coefficient of
the contribution that is logarithmic in the coupling constant,
\begin{align}
\int\frac{d^4p}{(2\pi)^4}2\pi\delta(p^2){\rm sign}(p^0)\frac{T^2}{2^6 \pi}
\left(\frac12+\frac{1}{{\rm e}^{-\beta p^0}-1}\right)
\frac{6\beta^3{\rm e}^{\beta p^0}}{\left({\rm e}^{\beta p^0}+1\right)^2}
=\frac{3}{2^{10}\pi}T\approx9.33\times10^{-4}T\,. 
\end{align}
The independence of the result (39) on $G$ and $k_*$ is valid up to order $G\log G$. The next-to leading expressions are of order $G^2 \log G$, such that for values of $G \sim 0.3$, one should expect to yield an accuracy of about 20\%, which is obviously less than what is suggested by the number of digits given in the numerical coefficients.
Note that this estimate
for the accuracy is very crude as it does not account for loop and
phase-space factors. An estimate of the next-to-leading order (NLO) contribution is non-trivial,
and a calculation of the NLO production rate of photons from the
quark-gluon plasma has recently been reported in
Ref.~\cite{Ghiglieri:2013gia}.

We can also extract the coefficients of the contributions that are linear and logarithmic in the couplings by directly performing
the integral~(\ref{C:wv:R}) for different values of the coupling. By this numerical fitting
procedure, we find
\begin{align}
\gamma^{{\rm fl(R,\ell)}\delta \ell} = 3.72\times 10^{-3} \times GT - 8.31\times 10^{-4} \times GT \log G\,. \label{eqn:fermi:num}
\end{align}
This decomposition is valid for a range of $G$ between $G=0.01$ and $G=0.6$ and can therefore also be used for the calculation of related processes in other Baryogenesis scenarios. The numerical difference between the results~(\ref{gamma:t-fermi:semi-an}) and~(\ref{eqn:fermi:num}) is due to
a partial inclusion of higher order effects that is implied when~(\ref{C:wv:R}) is integrated directly.

In Figure~\ref{fig:kSigma}, we show a comparison
of the result from the numerical
integration for ${\cal C}^{{\rm fl}{\rm (R)}}_\ell(p)$ and
the semi-analytic result Eq.~(\ref{C:ell:semi-analytical}).
There is a very good agreement for large $p^0/T$, while we
observe the anticipated breakdown of the approximations when
$p^0=|\mathbf p|\gg\sqrt{GT}$ is not valid.

We finally note that in the case of right-handed neutrino
production, there is no $t$-channel divergence from
the exchange of a left-handed SM-lepton $\ell$ when $M_N\not=0$.
Rather, this contribution has a logarithmic divergence
in $M_N^2$, as shown in Ref.~\cite{Garbrecht:2013gd},
indicating that the resummed propagator for $\ell$
should be used as well when $M_N\not=0$ but $M_N\ll T$.

\section{Vertex Type Contributions}\label{sec:vertex}

\begin{figure}[ht!]
\center
\epsfig{width=.4\textwidth, file=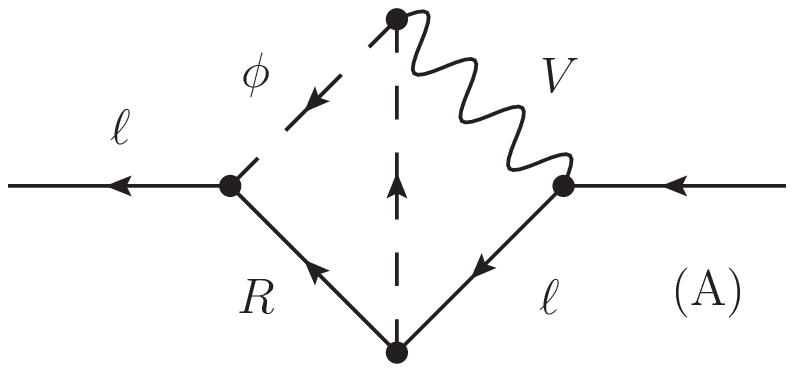}\hspace*{1.5cm}
\epsfig{width=.4\textwidth, file=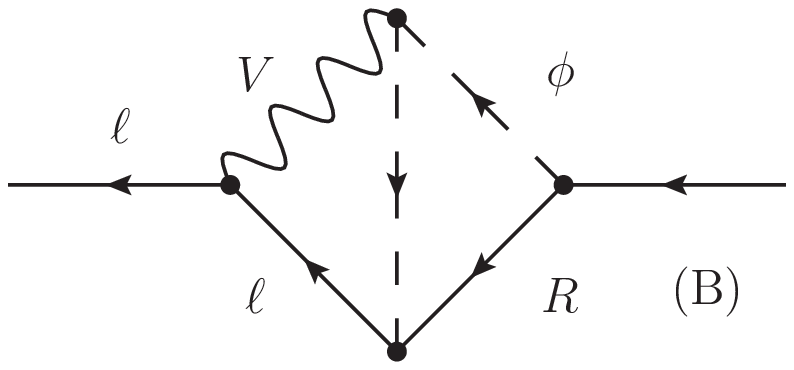}\\
\epsfig{width=.4\textwidth, file=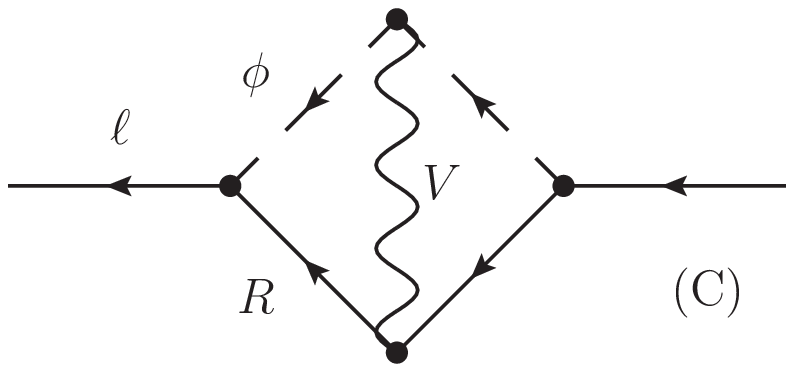}\\ 
\epsfig{width=.4\textwidth, file=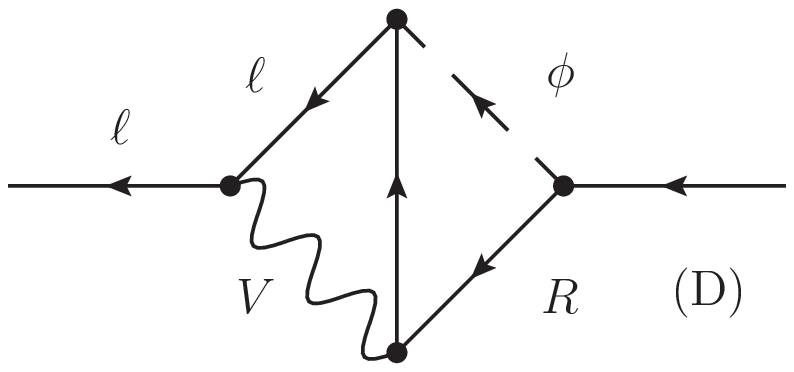}\hspace*{1.5cm}
\epsfig{width=.4\textwidth, file=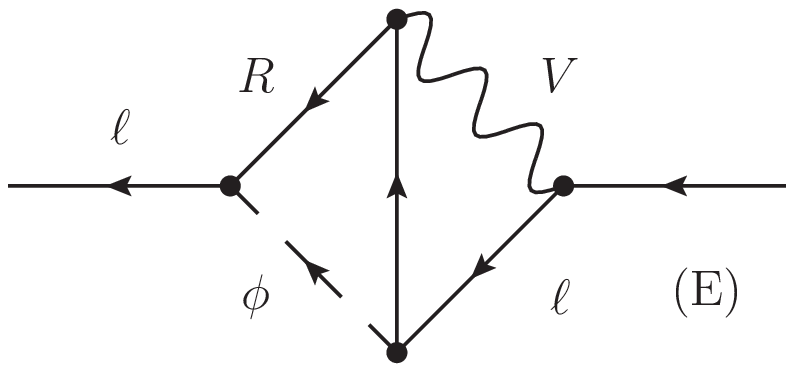}
\caption{
\label{fig:fl_vertex}
Two-loop self energies that contribute to the flavour relaxation
rate, where $V$ stands for both SU(2) and U(1) gauge bosons. Only (A) and (B) are allowed for SU(2) gauge bosons, but all five diagrams appear with U(1) gauge bosons. In the present approximation, we use tree-level propagators,
as indicated by the use of single lines, and 
the external legs are understood to be amputated.
}
\end{figure}

The two-loop self energies that involve two charged lepton Yukawa couplings and that descend from 2PI vacuum graphs are shown in Figure~\ref{fig:fl_vertex}. All diagrams are obtained from the one-loop self energy by connecting two different propagators with a gauge boson propagator. Note that diagrams~(C), (D), and~(E) only exist for the weak
hypercharge gauge boson and therefore are of order $g_1^2 h^2$. 

\begin{figure}[ht!]
\begin{center}
\epsfig{width=7cm, file=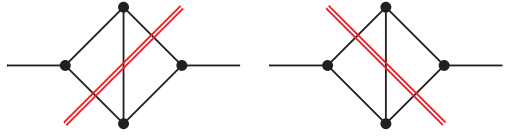}
\end{center}
\caption{
\label{fig:fl_vertex:cuts}
The cuts in the vertex diagrams (indicated by red double lines) that
correspond to scatterings. Note that the remaining possible cuts
correspond to virtual corrections to $1\leftrightarrow 2$ processes,
that are kinematically forbidden for the (approximately)
massless particles considered here.
}
\end{figure}

Since we consider massless particles, virtual corrections
to the vertices do not alter the fact that $1\leftrightarrow 2$ processes are
kinematically forbidden. Therefore, we can restrict the discussion to those
configurations that contribute to the $2\leftrightarrow 2$ scattering rates. Each diagram has two such contributions that are indicated by the cuts in Figure~\ref{fig:fl_vertex:cuts}.

In order to present our calculational method, we discuss in the following
the contributions that arise from diagram~(C).
The approach to calculating the
remaining diagrams is very similar and therefore presented more briefly.
As we do not need to use resummed propagators for the calculation of the
LO contributions from the vertex diagrams, all propagators that are explicitly employed in this Section are understood to be tree-level.
For notational simplicity, we omit the superscript $(0)$ that would
otherwise occur in a large number of instances.
The contribution to the CTP self energy represented in diagram~(C) is
\begin{align}
{\rm i}\slashed\Sigma_\ell^{{\rm (C)}ab}=&Y_{\rm R} Y_\phi g_1^2 h^\dagger
\int \frac{d^4 k}{(2\pi)^4}
\int \frac{d^4 q}{(2\pi)^4}
\sum\limits_{c,d}cd\,
{\rm i}S_{\rm R}^{ac}(k)
\gamma^\mu{\rm i}\Delta_{\mu\nu}^{cd}(k+q-p) [k-p-q]^\nu
\notag\\
&\hskip4cm
\times {\rm i}S_{\rm R}^{cb}(p-q)
{\rm i}\Delta_\phi^{db}(q)
{\rm i}\Delta_\phi^{ad}(p-k)
h\,,
\end{align}
where $Y_{\rm R}=-1$ and $Y_\phi=\frac12$. The collision term only depends on ${\rm i} \Sigma\!\!\!/^{<,>}$. We therefore consider
\begin{align}
\notag
{\rm i}\slashed\Sigma_\ell^{{\rm (C)}>}(p)=&Y_{\rm R} Y_\phi g_1^2\, h^\dagger
\int\frac{d^4k}{(2\pi)^4}\int\frac{d^4q}{(2\pi)^4}
\\\notag
\bigg\{&
{\rm i}S_{\rm R}^{--}(k)\gamma^\mu
{\rm i} S_{\rm R}^{-+}(p-q)
{\rm i}\Delta_\phi^{-+}(q)[k-p-q]^\nu
{\rm i}\Delta_\phi^{--}(p-k)
{\rm i}\Delta^{--}_{\mu\nu}(k+q-p)
\\\notag
-&
{\rm i}S_{\rm R}^{-+}(k)\gamma^\mu
{\rm i} S_{\rm R}^{++}(p-q)
{\rm i}\Delta_\phi^{-+}(q)[k-p-q]^\nu
{\rm i}\Delta_\phi^{--}(p-k)
{\rm i}\Delta^{+-}_{\mu\nu}(k+q-p)
\\\notag
-&
{\rm i}S_{\rm R}^{--}(k)\gamma^\mu
{\rm i} S_{\rm R}^{-+}(p-q)
{\rm i}\Delta_\phi^{++}(q)[k-p-q]^\nu
{\rm i}\Delta_\phi^{-+}(p-k)
{\rm i}\Delta^{-+}_{\mu\nu}(k+q-p)
\\\notag
+&
{\rm i}S_{\rm R}^{-+}(k)\gamma^\mu
{\rm i} S_{\rm R}^{++}(p-q)
{\rm i}\Delta_\phi^{++}(q)[k-p-q]^\nu
{\rm i}\Delta_\phi^{-+}(p-k)
{\rm i}\Delta^{++}_{\mu\nu}(k+q-p)
\bigg\}
\\
\times & h
\,.
\end{align}
It is useful to shift the momenta in the first and the third term,
such that
\begin{align}
\notag
{\rm i}\slashed\Sigma_\ell^{{\rm (C)}>}(p)=&Y_{\rm R} Y_\phi g_1^2 h^\dagger
\int\frac{d^4k}{(2\pi)^4}\int\frac{d^4q}{(2\pi)^4}
\\\notag
\bigg\{&
{\rm i}S_{\rm R}^{--}(p-q)\gamma^\mu
{\rm i} S_{\rm R}^{-+}(k)
{\rm i}\Delta_\phi^{-+}(p-k)[k-p-q]^\nu
{\rm i}\Delta_\phi^{--}(q)
{\rm i}\Delta^{--}_{\mu\nu}(k+q-p)
\\\notag
-&
{\rm i}S_{\rm R}^{-+}(k)\gamma^\mu
{\rm i} S_{\rm R}^{++}(p-q)
{\rm i}\Delta_\phi^{-+}(q)[k-p-q]^\nu
{\rm i}\Delta_\phi^{--}(p-k)
{\rm i}\Delta^{+-}_{\mu\nu}(k+q-p)
\\\notag
-&
{\rm i}S_{\rm R}^{--}(p-q)\gamma^\mu
{\rm i} S_{\rm R}^{-+}(k)
{\rm i}\Delta_\phi^{++}(p-k)[k-p-q]^\nu
{\rm i}\Delta_\phi^{-+}(q)
{\rm i}\Delta^{+-}_{\mu\nu}(k+q-p)
\\\notag
+&
{\rm i}S_{\rm R}^{-+}(k)\gamma^\mu
{\rm i} S_{\rm R}^{++}(p-q)
{\rm i}\Delta_\phi^{++}(q)[k-p-q]^\nu
{\rm i}\Delta_\phi^{-+}(p-k)
{\rm i}\Delta^{++}_{\mu\nu}(k+q-p)
\bigg\}
\\
\times&h
\,.
\end{align}
In the collision term, the Dirac structure of $i\Sigma\!\!\!/_\ell^{<,>}(p)$ is dotted into the lepton propagator $iS_\ell^{>,<}(p)$, which provides a factor $p\!\!\!/$. Taking the trace over Dirac indices and using that ${\rm tr}[\gamma^\mu \gamma^\nu \gamma^\rho \gamma^\sigma] = {\rm tr}[\gamma^\sigma \gamma^\rho \gamma^\nu\gamma^\mu]$, as well as the cyclicity of the trace, we arrive at the form
\begin{align}
\notag
\label{SigGreaterC}
{\rm i}\,{\rm tr}&[\slashed p \slashed\Sigma_\ell^{{\rm (C)}>}(p)]=
Y_{\rm R} Y_\phi g_1^2 h^\dagger
\int\frac{d^4k}{(2\pi)^4}\int\frac{d^4q}{(2\pi)^4}
{\rm tr}\bigg[\slashed p
\\\notag
\times\bigg\{&
{\rm i}S_{\rm R}^{--}(p-q)\gamma^\mu
{\rm i} S_{\rm R}^{-+}(k)
{\rm i}\Delta_\phi^{-+}(p-k)[k-p-q]^\nu
{\rm i}\Delta_\phi^{--}(q)
{\rm i}\Delta^{A--}_{\mu\nu}(k+q-p)
\\\notag
-&
{\rm i}S_{\rm R}^{++}(p-q)\gamma^\mu
{\rm i} S_{\rm R}^{-+}(k)
{\rm i}\Delta_\phi^{--}(p-k)[k-p-q]^\nu
{\rm i}\Delta_\phi^{-+}(q)
{\rm i}\Delta^{A+-}_{\mu\nu}(k+q-p)
\\\notag
-&
{\rm i}S_{\rm R}^{--}(p-q)\gamma^\mu
{\rm i} S_{\rm R}^{-+}(k)
{\rm i}\Delta_\phi^{++}(p-k)[k-p-q]^\nu
{\rm i}\Delta_\phi^{-+}(q)
{\rm i}\Delta^{A+-}_{\mu\nu}(k+q-p)
\\\notag
+&{\rm i}S_{\rm R}^{++}(p-q)\gamma^\mu
{\rm i} S_{\rm R}^{-+}(k)
{\rm i}\Delta_\phi^{-+}(p-k)[k-p-q]^\nu
{\rm i}\Delta_\phi^{++}(q)
{\rm i}\Delta^{A++}_{\mu\nu}(k+q-p)
\bigg\}
\bigg]
\\
\times& h
\,.
\end{align}
First note that it is not possible to put all five propagators
simultaneously on shell. Now suppose that $p-q$ is off shell,
whereas the remaining momenta are on shell. Then, the second and
third term cancel, as well as the first and the fourth. Note that
this corresponds to the kinematic regime of the $CP$-violating
source of the corresponding diagram for Leptogenesis~\cite{flbg},
where the
cancellation does not occur in general, because of the complex
conjugation of coupling constants.

When three momenta are on shell, the first and the fourth term
give a virtual correction to the kinematically
suppressed $1\leftrightarrow 2$ process.
The $2\leftrightarrow 2$ scatterings are encapsulated within the
second and the third term, where the two off-shell momenta
are $p-q$ and $k-p$.
The numerator algebra of the self energy~(\ref{SigGreaterC})
is given by
\begin{align}
{\rm tr}
[\slashed{p}(\slashed{p}-\slashed{q})(\slashed{k}-\slashed{p}-\slashed{q})\slashed{k}]
=8p\cdot k\,p\cdot q-4p^2 k\cdot(p+2q)\,.
\end{align}
Here, we have used the on-shell conditions $k^2=q^2=(p-k-q)^2=0$,
that apply for the second and the third term, while we yet allow for
$p^2\not=0$, such that this result may be used for the application
of calculating the production rate of massive
right-handed neutrinos~\cite{Garbrecht:2013gd}. In what follows however,
we set $p^2=0$. This leads to the great simplification that the
zeros in the numerator and denominator cancel:
\begin{align}
\frac12{\rm tr}
[\slashed{p}(\slashed{p}-\slashed{q})(\slashed{k}-\slashed{p}-\slashed{q})\slashed{k}]
\underset{p^2=0}{=}(p-q)^2(k-p)^2\,.
\end{align}
A consequence of the fact that this cancellation does not occur for
$p^2\not=0$ is the presence of soft and collinear divergences in the
particular real and virtual contributions to the vertex-type self energy
for the production of massive neutrinos. It is shown in
Refs.~\cite{Salvio:2011sf,Laine:2011pq,Garbrecht:2013gd}
that upon summation of all contributions, these soft and collinear 
divergences cancel.

The same discussion can be repeated for $\slashed \Sigma^{{\rm (C)}<}_\ell$
by simply exchanging the first and the second CTP index on all propagators. For the integrated collision term, we then obtain
\begin{subequations}
\begin{align}
\int \frac{d^4 p}{(2 \pi)^4} {\cal C}^{\rm fl (C)}_\ell
=& -Y_{\rm R} Y_\phi g_1^2\int\frac{d^4p}{(2\pi)^4}\int\frac{d^4k}{(2\pi)^4}\int\frac{d^4q}{(2\pi)^4}
2\pi\delta(p^2)2\pi\delta(k^2)2\pi\delta(q^2)
\label{eqn:flC}
\\\notag
\times& 2\pi\delta((k+q-p)^2)
{\rm sign}(p^0)\,{\rm sign}(k^0)\,{\rm sign}(q^0)\,{\rm sign}(k^0+q^0-p^0)
\notag\\\notag
\times&
{\cal G}(q^0,k^0+q^0-p^0,k^0,p^0) \,,
\\
{\cal G}(E_1,E_2&,E_3,E_4)=
(1+f_\phi(E_1))f_A(E_2)h^\dagger (1- f_{\rm R}(E_3)) h f_{\ell}(E_4)
\\\notag
&\qquad\qquad-f_\phi(E_1)(1+f_A(E_2))h^\dagger f_{\rm R}(E_3) h (1-f_{\ell}(E_4))\,.
\end{align}
%
Using the definition~(\ref{eqn:coll:ell}), we obtain the
flavour-sensitive rate
\begin{align}
\Gamma_\ell^{{\rm fl}{\rm (C)}}
=& Y_{\rm R} Y_\phi g_1^2\int\frac{d^4p}{(2\pi)^4}\int\frac{d^4k}{(2\pi)^4}\int\frac{d^4q}{(2\pi)^4}
2\pi\delta(p^2)2\pi\delta(k^2)2\pi\delta(q^2)2\pi\delta((k+q-p)^2)
\notag\\
\times&
{\rm sign}(p^0)\,{\rm sign}(k^0)\,{\rm sign}(q^0)\,{\rm sign}(k^0+q^0-p^0)
\notag\\
\times& \frac{1}{2} \left( 
{\cal G}(q^0,k^0+q^0-p^0,k^0,p^0) + {\rm h.c.} \right),
\end{align}
\end{subequations}
where the hermitian conjugation acts on the implicit
flavour indices in ${\cal G}$. 
In order to extract the coefficients
$\gamma^{{\rm fl}{\rm (C)}\delta\ell,\delta{\rm R}}$, we linearise in the chemical potentials for $\ell$ and ${\rm R}$. Since
${\cal G}$ is odd under a simultaneous exchange of
$E_{1,3}\leftrightarrow E_{2,4}$ and $\phi,\ell\leftrightarrow A,{\rm R}$,
we can do this calculation for $\mu_\ell$ and directly infer
the result for $\mu_{\rm R}$. To linear order in $\mu_\ell$,
\begin{subequations}
\begin{align}
{\cal G}
=&
\frac{{\rm e}^{\beta E_1+\beta E_3}+{\rm e}^{\beta E_2}}
{({\rm e}^{\beta E_1}-1)({\rm e}^{\beta E_2}-1)({\rm e}^{\beta E_3}+1)}
\frac{{\rm e}^{\beta E_4}}{({\rm e}^{\beta E_4}+1)^2}\beta
h^\dagger h \mu_\ell
=\bar{\cal G} h^\dagger h q_\ell
\,,
\\
\mu_\ell=&6\beta^2 q_\ell\,.
\end{align}
\end{subequations}
Using these relations, we can extract the reduced interaction rates: 
\begin{align} 
\gamma^{{\rm fl(C)}\delta \ell}=&-\gamma^{{\rm fl(C)}\delta {\rm R}} \label{eqn:rates:C}
\\\notag
=&
 Y_{\rm R} Y_\phi
g_1^2\int\frac{d^4p}{(2\pi)^4}\int\frac{d^4k}{(2\pi)^4}\int\frac{d^4q}{(2\pi)^4}
2\pi\delta(p^2)2\pi\delta(k^2)2\pi\delta(q^2)2\pi\delta((p-k-q)^2)
\\\notag
&\times
{\rm sign}(p^0)\,{\rm sign}(k^0)\,{\rm sign}(q^0)\,{\rm sign}(k^0+q^0-p^0)
\bar {\cal G}(q^0,k^0+q^0-p^0,k^0,p^0)
\\\notag
\equiv& Y_{\rm R} Y_\phi g_1^2 \, \gamma^{{\rm fl}0}_{\rm vert}
\\\notag
=& 
7.72\times 10^{-4} Y_{\rm R} Y_\phi g_1^2 T
\,,
\end{align}
where we have defined $\gamma_{\rm vert}^{\rm fl 0}$, the universal value of the phase space integrals in~(\ref{eqn:rates:C}). 

Now, for diagrams~(A), (B) and (D), (E), it is possible to put three propagators on shell without cutting through the gauge boson propagator. Those cuts correspond to the interference between a Higgs and a gauge 
boson mediated process, and do not contribute to the equilibration of flavours. Indeed, after linearising in deviations from equilibrium, these terms cancel in the collision term.

Following the calculation of diagram~(C),
the relevant part of the self energy diagram~(A) then evaluates to
(still, we suppress the superscript $(0)$ on the tree-level propagators) 
\begin{align}
	{\rm i} \SigSlash^{(A)>}_\ell(p) & = - \left({\textstyle \frac{3}{4}} g_2^2 + Y_\ell Y_\phi g_1^2 \right) h^\dagger 
	\int  \frac{d^4 k}{(2\pi)^4} \frac{d^4 q}{(2\pi)^4} \times \notag \\
	&  {\rm i} S^>_{\rm R} (k) {\rm i} S^{T}_\ell(p-q) \left[ 2 p\!\!\!/ - 2 k\!\!\!/ -q\!\!\!/ \right] {\rm i} \bar\Delta_A^>(q) {\rm i} \Delta_\phi^{\bar{T}}(p-k) {\rm i} \Delta_\phi^>(p-q-k) \times h\,,
\end{align}
where we used that ${\rm i}\Delta_{A\mu\nu}=-g_{\mu\nu}{\rm i}\bar\Delta_A$. To facilitate the comparison with the result for (C), we can use that in thermal equilibrium (with vanishing chemical potentials),
${\rm i}\Delta^>(k) = {\rm i} \Delta^<(-k)$, and  ${\rm i} \bar\Delta_A(k) = {\rm i} \Delta_\phi(k)$. 
Inserting this into the collision term and taking the trace over Dirac indices, we find
\begin{align}
	{\cal{C}}_\ell^{\rm fl(A)} & = \left({\textstyle \frac{3}{4}} g_2^2 + Y_\ell Y_\phi g_1^2 \right) 
	\int  \frac{d^4 k}{(2\pi)^4} \frac{d^4 q}{(2\pi)^4}
	2\pi\delta(p^2)2\pi\delta(k^2)2\pi\delta(q^2)
\\\notag
\times& 2\pi\delta((k+q-p)^2)
{\rm sign}(p^0)\,{\rm sign}(k^0)\,{\rm sign}(q^0)\,{\rm sign}(k^0+q^0-p^0)
\notag\\\notag
\times&
{\cal G}(q^0,k^0+q^0-p^0,k^0,p^0)\,,
\end{align}
where we have used that 
\begin{align}
{\rm tr}\left[ P_R (2 \slashed{p} - 2 \slashed{k} - \slashed{q} ) 
	(\slashed{p} - \slashed{q}) \slashed{k} \slashed{p} \right]
&=(p-q)^2(p-k)^2\;\;\textnormal{for}\;\;
(p-k-q)^2=k^2=p^2=q^2=0\,,
\end{align}
which again cancels with the denominators of the off-shell propagators. 
The contribution from diagram (B) can easily be seen to be identical to that of diagram (A). Comparing with Eq.~(\ref{eqn:flC}), it follows that 
\begin{align}
	\gamma^{\rm fl (A) \delta \ell } + \gamma^{\rm fl(B) \delta \ell} = -\gamma^{\rm fl (A) \delta R } - \gamma^{\rm fl(B) \delta R}  = - \left({\textstyle\frac{3}{2}} g_2^2 + 2 Y_\ell Y_\phi g_1^2 \right) \gamma^{\rm fl 0}_{\rm vert}\,.
\end{align}
Another way of verifying that these contributions come with the same numerical coefficient is to note
that diagrams~(C) and (A), (B) descend
from three-loop vacuum diagrams, that are identical in terms of spin $0$, $1/2$
and $1$ propagators (up to exchanges of $\ell$ and ${\rm R}$ and of
the ${\rm SU}(2)_{\rm L}$ and ${\rm U}(1)_Y$ gauge bosons). These vacuum
diagrams are (up to the $d^3 p$ integration) recovered through the
integration over $d p^0$.

The relevant contribution from diagram~(D) to the self energy
is given by
\begin{align}
{\rm i}\slashed\Sigma^{{\rm (D)}>}=&
-Y_{\ell} Y_{\rm R} g_1^2 h^\dagger\int\frac{d^4 q}{(2\pi)^4}\frac{d^4 k}{(2\pi)^4}
\\\notag
\times&
\gamma^\mu {\rm i}S_\ell^{\bar T}(p-k){\rm i}S_{\rm R}^>(p-k-q)\gamma_\mu 
{\rm i} S_{\rm R}^T(p-q){\rm i}\bar\Delta^>(k)
{\rm i}\Delta_\phi^<(-q)  h
\,.
\end{align}
The relevant cut through diagram (E) can be brought into the same form.
When inserting this into the collision term, we obtain
\begin{align}
{\cal C}^{{\rm fl}{\rm(D,E)}}=&
-Y_{\ell} Y_{\rm R} g_1^2 h^\dagger\int\frac{d^4 q}{(2\pi)^4}\frac{d^4 k}{(2\pi)^4}
\frac{1}{(p-k)^2}\frac{1}{(p-q)^2}
\\\notag
\times&
{\rm tr}[P_{\rm R}\gamma^\mu(\slashed p-\slashed k)(\slashed p-\slashed k-\slashed q)
\gamma_\mu(\slashed p -\slashed q)\slashed p]
\\\notag
\times&
\left[ {\rm i}\bar S^>_{\rm R}(p-k-q){\rm i}\bar\Delta_A^>(k){\rm i}\Delta^>_\phi(q)  h {\rm i}\bar{S}_\ell^<(p) - (>\leftrightarrow <) \right] \,.
\end{align}
The Dirac trace is now different than for the diagrams~(A), (B) and~(C),
but with the on-shell conditions, it reduces to
\begin{align}
{\rm tr}[P_{\rm R}\gamma^\mu(\slashed p-\slashed k)(\slashed p-\slashed k-\slashed q)
\gamma_\mu(\slashed p -\slashed q)\slashed p]=
2(p-k)^2(p-q)^2\,,
\end{align}
such that the collision term simplifies to 
\begin{align}
{\cal C}^{{\rm fl}{\rm(D,E)}}=&
-2 Y_{\ell} Y_{\rm R} g_1^2 h^\dagger\int\frac{d^4 q}{(2\pi)^4}\frac{d^4 k}{(2\pi)^4}
\left[{\rm i}\bar S^>_{\rm R}(p-k-q){\rm i}\bar\Delta_A^>(k){\rm i}\Delta^>_\phi(q)  h {\rm i}\bar{S}_\ell^<(p) - (>\leftrightarrow <) \right]
 \\\notag
=&
-2 Y_{\ell} Y_{\rm R} g_1^2 h^\dagger\int\frac{d^4 q}{(2\pi)^4}\frac{d^4 k}{(2\pi)^4}
\left[{\rm i}\bar S^>_{\rm R}(k){\rm i}\bar\Delta_A^<(k+q-p){\rm i}\Delta^>_\phi(q)  h{\rm i}\bar{S}_\ell^<(p) - (>\leftrightarrow <) \right],
\end{align}
where in the last term, we have shifted the momentum $k \to p-k-q$, and ${\rm i}\bar{S}$ is defined by ${\rm i} S(p) = {\rm i} \slashed p \bar{S}(p)$. Comparing with the calculations for (C) and (A,B), it is then easy to verify that 

\begin{align}
\gamma^{{\rm fl(D)}\delta \ell}+\gamma^{{\rm fl(E)}\delta \ell}
=&-\gamma^{{\rm fl(D)}\delta {\rm R}}-\gamma^{{\rm fl(E)}\delta {\rm R}}
=-4 Y_{\rm R}Y_{\ell} g_1^2 \gamma^{{\rm fl}0}_{\rm vert}
\,.
\end{align}
As the sum of the contributions from the various diagrams, we obtain
\begin{align}
\gamma^{{\rm fl}{\rm (A+B+C+D+E)}\delta\ell}
=&-\gamma^{{\rm fl}{\rm (A+B+C+D+E)}\delta{\rm R}}
=\gamma^{\rm fl}_{\rm vertex}
\\\notag
=&
\left[
-\frac 32 g_2^2+
(2Y_{\rm R}Y_\phi
-2Y_{\rm L}Y_\phi
-4Y_{\rm L} Y_{\rm R})g_1^2
\right]
\gamma^{{\rm fl}0}_{\rm vert}\,.
\end{align}
Finally we can insert the weak hypercharges in order to obtain the flavour relaxation rate, and find
\begin{align}
	\gamma^{\rm fl}_{\rm vertex} &= - \left( {\textstyle \frac 32} g_2^2+  {\textstyle \frac 52} g_1^2 \right) \times 7.72 \times 10^{-4} \times T  \\\notag
	& = -7.72 \times 10^{-4} \times GT -  2 \times 7.72 \times 10^{-4} \times g_1^2 T\,.
\end{align}

\section{$1\leftrightarrow2$ Processes}\label{sec:onetwo}

The production rate of the singlet neutrinos also includes
$1\leftrightarrow 2$ processes~\cite{Anisimov:2010gy,Besak:2012qm}.
In the limit where
$M_N \gg T$, the processes $\ell \phi \to N$ and $\bar\ell \phi^*\to N$
are the main contributions to the production rate. At higher
temperatures relative to $M_N$,
which is relevant for the weak washout regime, the thermal masses
of the lepton and the Higgs bosons,
\begin{align}
m_\ell^2&=\frac1{16} (3g_2^2 + g_1^2)T^2\,,
\\
m_\phi^2&=\frac1{16}(3g_2^2+ g_1^2+4 h_t^2 +8 \lambda) T^2\,,
\end{align}
are of importance. These masses are understood to be effective masses
valid for modes of momenta larger than $g T$.

Following Eqn.~(\ref{eqn:coll:n2}),
the tree level rates can be obtained from
\begin{align}
\label{SigmaA:tree}
{\rm tr}[\slashed p\Sigma^{\cal A}(p)]
=\frac1{4\pi}|Y|^2\frac{|M_N^2+m_\ell^2-m_\phi^2|}{|\mathbf p|}
\left[{\cal I}_{\ell}(\omega_{\ell+})-{\cal I}_{\ell}(\omega_{\ell-}))\right]\,,
\end{align}
where
\begin{align}
\label{limits:omegaell}
\omega_{\ell\pm}=&
\frac{|p^0|}{2M_N^2}\left|M_N^2+m_\ell^2-m_\phi^2\right|
\\\notag
\pm&\frac{1}{2M_N^2}
\sqrt{
\left({p^0}^2-M_N^2\right)
\left(
M_N^4+m_\ell^4+m_\phi^4-2M_N^2m_\ell^2-2m_\ell^2m_\phi^2-2M_N^2m_\phi^2
\right)}\,.
\end{align}
We take the singlet neutrino to be on shell, $p^2=M_N^2$, and
\begin{align}
{\cal I}_\ell(\omega_\ell)=\left\{
\begin{array}{ll}
-\omega_\ell
-\frac1\beta\log\left({\rm e}^{\beta(p^0-\omega_\ell)}-1\right)
+\frac1\beta\log\left({\rm e}^{\beta \omega_\ell}+1\right) \;
&
\textnormal{for}\;\;
\begin{array}{l}
M_N>m_\ell+m_\phi\\
\textnormal{and}\;\;
m_\ell>M_N+m_\phi
\end{array}
\\
\frac1\beta\log\left({\rm e}^{\beta(p^0+\omega_\ell)}-1\right)
-\frac1\beta\log\left({\rm e}^{\beta \omega_\ell}+1\right)
&
\textnormal{for}\;\;m_\phi>M_N+m_\ell
\\
0 & \textnormal{otherwise}
\end{array}
\right.
\,.
\end{align}

However, there are also processes involving the multiple collinear
emission of soft gauge bosons from the lepton and the Higgs
boson propagators. While kinematically not possible in the vacuum
(at LO, when all scattering particles are massless),
here they occur because of the thermal masses of the
gauge bosons. Note that in the scattering diagrams calculated
in the preceding Sections, we have
approximated the gauge boson masses by {\it zero}, which means that
the collinear processes are not readily included.

The summation of the collinear processes is derived and
discussed in
Ref.~\cite{Anisimov:2010gy}. Here, we just quote the procedure.
First, we solve the integral equations
\begin{subequations}
\label{integral:equations}
\begin{align}
{\rm i}\varepsilon(k_\parallel,{\mathbf p})\bm{f}(\mathbf p_\perp,p_\parallel,k_\parallel)
-\int\frac{d^2q_\perp}{(2\pi)^2}
{\cal C}(\mathbf q_\perp)
\left[
\bm{f}(\mathbf p_\perp,p_\parallel,k_\parallel)-
\bm{f}(\mathbf p_\perp-\mathbf q_\perp,p_\parallel,k_\parallel)
\right]
&=2\mathbf p_\perp\,,
\\
{\rm i}\varepsilon(k_\parallel,{\mathbf p})\psi(\mathbf p_\perp,p_\parallel,k_\parallel)
-\int\frac{d^2q_\perp}{(2\pi)^2}
{\cal C}(\mathbf q_\perp)
\left[
\psi(\mathbf p_\perp,p_\parallel,k_\parallel)-
\psi(\mathbf p_\perp-\mathbf q_\perp,p_\parallel,k_\parallel)
\right]
&=1\,,
\end{align}
\end{subequations}
where
\begin{align}
\varepsilon(k_\parallel,\mathbf p)=
\frac{k_\parallel}{2p_\parallel(p_\parallel-k_\parallel)}
\left(
\mathbf p_\perp^2
+\frac{
p_\parallel(p_\parallel-k_\parallel)M_N^2
-k_\parallel(p_\parallel-k_\parallel)m_\ell^2
-k_\parallel p_\parallel m_\phi^2
}
{k_\parallel^2}
\right)
\end{align}
and
\begin{align}
{\cal C}(\mathbf q_\perp)=
\frac1\beta\left[
\frac34 g_2^2
\left(
\frac1{\mathbf q_\perp^2}
-\frac1{\mathbf q_\perp^2+m_{{\rm D}2}^2}
\right)
+\frac12g_1^2
\left(
\frac1{\mathbf q_\perp^2}
-\frac1{\mathbf q_\perp^2+m_{{\rm D}Y}^2}
\right)
\right]
\,.
\end{align}
The Debye masses are
$m_{{\rm D}2}=\frac{11}6g_2^2 T^2$ and
$m_{{\rm D}Y}=\frac{11}6g_1^2 T^2$.
The solution to Eqs.~(\ref{integral:equations}) is
best performed in impact parameter space
and it is not straightforward. We refer to Ref.~\cite{Anisimov:2010gy}
for the details.

Then, the production rate for singlet neutrinos is given by
\begin{align}
{\rm tr}\left[\slashed k {\rm i}\slashed\Sigma^<(k)\right]
=-2 Y^2 \frac{k^0}{|\mathbf k|}
\int\frac{d^3 p}{(2\pi)^3}&\frac{1}{|\mathbf k|-p_\parallel}
\frac{1}{{\rm e}^{\beta p_\parallel}+1}
\frac{1}{{\rm e}^{\beta(|\mathbf k|-p_\parallel)}-1}
\\\notag
\times&
{\rm Re}\left[
\frac{|\mathbf k|}{2 p_\parallel}
\mathbf p_\perp\cdot\bm{f}(\mathbf p_\perp,p_\parallel,|\mathbf k|)
+\frac{M_N^2}{|\mathbf k|}\psi(\mathbf p_\perp,p_\parallel,|\mathbf k|)
\right]
\,,
\end{align}
where we take $k^0>0$.
The integration over $d^3 p$ is performed in the limits
\begin{align}
p_{\parallel\pm}
=&\frac{|\mathbf k|}{2M_N^2}\bigg[
(M_N^2+m_\ell^2-m_\phi^2)
\\\notag
\pm&
\sqrt{
M_N^4+m_\ell^4+m_\phi^4-2M_N^2m_\ell^2-2m_\ell^2m_\phi^2-2M_N^2m_\phi^2
}
\bigg]\,,
\end{align}
while the integration over $|\mathbf p_\perp|$ is
subsequently performed from {\it zero} to infinity.

An analytic calculation or approximation of the
$1\leftrightarrow 2$ rates contributing to
the production of $N$ is presently not available, and the numerical
evaluation following above procedure is yet time-consuming.
Moreover, a numerical generalisation to the situation where also
the external fermion line can radiate a gauge boson has not yet
been performed.
In order to obtain
an estimate of the uncertainty in the flavour relaxation
rate due to the radiation from the external
fermion line, but also in order to investigate the effect of
the scale ({\it i.e.} temperature) dependence of the gauge
coupling constants on the right-handed neutrino production rate,
we calculate $\Gamma_N$ for varying values of $g_1$ and $g_2$.

\begin{figure}[t!]
\begin{center}
\epsfig{file=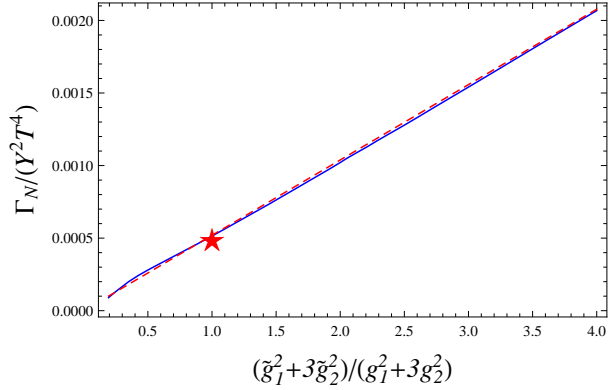,width=8cm}
\end{center}
\caption{
\label{figure:GaVarg1g2}
Right-handed neutrino production rate $\Gamma_N/(Y^2 T^4)$ for
$g_{1,2}\to\tilde g_{1,2}$, $M_N=0$, varying $\tilde g_{1,2}$,
and all remaining couplings as given in
Table~\ref{tab:couplings} for the scale $10^9{\rm GeV}$ (solid blue line).
We also indicate the fit~(\ref{1to2:fit}) (dashed red line) and
indicate the rate in the SM (red star).
}
\end{figure}

\begin{figure}[t!]
\begin{center}
\epsfig{file=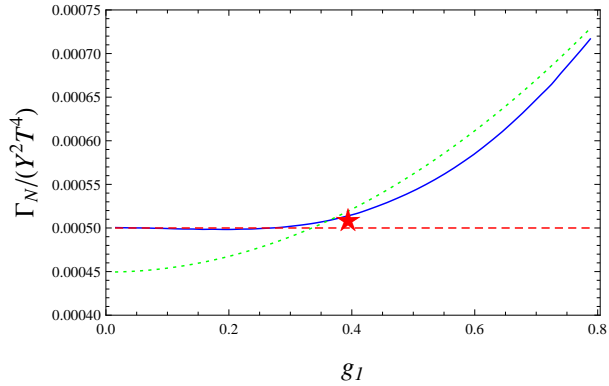,width=8cm}
\end{center}
\caption{
\label{figure:GaVarg1}
Right handed neutrino production rate $\Gamma_N/(Y^2T^4)$ for
$M_N=0$,
varying $g_1$ and all remaining couplings as given in
Table~\ref{tab:couplings} for the scale $10^9\,{\rm GeV}$ (solid blue line).
The dashed red line indicates the result for $g_1=0$, the dotted
green line the fit~(\ref{1to2:fit}).
The rate in the SM is indicated by a red star.
}
\end{figure}

The numerical results are presented in Figures~\ref{figure:GaVarg1g2}
and~\ref{figure:GaVarg1}. As in this work, we are interested in the
production of light (relativistic) right-handed neutrinos and
the scatterings of massless Standard Model leptons, gauge- and Higgs-bosons, we take for the mass $M_N$ of the right-handed neutrino
$M_N=0$. For the Standard Model couplings implied by
a Higgs boson mass of $125\,{\rm GeV}$, we thereby reproduce the
value $\Gamma_N^{1\leftrightarrow 2}\approx 5\times 10^{-4}/(Y^2T^4)$
found in Ref.~\cite{Anisimov:2010gy}. Moreover,
we find that
a good fit to the behaviour apparent in Figure~\ref{figure:GaVarg1g2}
is provided by
\begin{align}
\label{1to2:fit}
\Gamma_N^{1\leftrightarrow2} / (Y^2 T^4) \approx 8.8\times 10^{-4} G\,.
\end{align}
The results presented in Figure~\ref{figure:GaVarg1} indicate that
this formula becomes less accurate when one of the couplings is very
small, {\it i.e.} we observe a deviation from the
relation~(\ref{1to2:fit}) when $g_1\ll g_2$.

Regarding the flavour equilibration rate $\gamma^{\rm fl}$, we note
that the technique developed in Ref.~\cite{Anisimov:2010gy}
applies to diagrams of the type of Figure~\ref{fig:fl_vertex}(C),
where the gauge radiation originates from internal lines of the diagram
in Figure~\ref{fig:diagrams} only,
but not directly to the diagrams in Figures~\ref{fig:fl_vertex}(A,B,D,E),
where the gauge radiation also attaches to an external line.
A generalisation of the methods developed in Ref.~\cite{Anisimov:2010gy}
is beyond the scope of the present work, but as we see below, the
error from neglecting gauge radiation from an external line is quantitatively
small compared to the dominating contribution from the $t$-channel
exchange of fermions and the expected NLO corrections.
In order to make an estimate for the relaxation rate of left-handed
flavour $\gamma^{\rm fl}$, we notice that this should be complementary
to the relaxation rate $\gamma_R^{\rm fl}$
for right-handed flavour,
\begin{align}
\gamma^{\rm fl}=\frac12\gamma_R^{\rm fl}\,,
\end{align}
where the factor $1/2$ is due to the multiplicity of left-handed
leptons and Higgs bosons. Now, the right handed leptons ${\rm R}$ are different
from the right-handed singlet neutrinos $N$ in that they have the weak
hypercharge $-1$ and that within their self-energy diagram, no
charge-conjugated particles are running.
We may therefore approximate
\begin{align}
{\rm tr}[\slashed p\Sigma_R^{1\leftrightarrow 2}(p)]\approx\frac12{\rm tr}[\slashed p\Sigma_N^{1\leftrightarrow 2}(p)]\,,
\end{align}
and from Figures~\ref{figure:GaVarg1g2} and~\ref{figure:GaVarg1}, we
can estimate the relative uncertainty due to inaccurately neglecting
the ${\rm U}(1)$ weak hypercharge interactions by 15\%.
As it turns out that the $1\leftrightarrow 2$ rates are small compared
to the sum of the $2\leftrightarrow 2$ rates (which are in turn dominated
by the $t$-channel fermion exchange), and in view of the uncertainty
from neglecting contributions $\propto G^2 \log G^{-1}$ at the following 
order (NLO), it is therefore quantitatively sufficient to make the estimate
\begin{align}
{\rm tr}[\slashed p\Sigma_\ell^{1\leftrightarrow 2}(p)]\approx\frac14{\rm tr}[\slashed p\Sigma_N^{1\leftrightarrow 2}(p)]\,.
\end{align}
Substitution into Eqs.~(\ref{eqn:coll:ell}) and~(\ref{fl:rate})
and numerical evaluation then
yields (for the values of the couplings given in
Table~\ref{tab:couplings} at
the scale of $10^9 {\rm GeV}$)
\begin{align}
\gamma^{\rm fl}_{1\leftrightarrow 2}=9.9\times 10^{-4} T
\,.
\end{align}
When expressed as a fit similar to relation~(\ref{1to2:fit}), this becomes
\begin{align}
\gamma^{\rm fl}_{1\leftrightarrow 2}=1.7\times 10^{-3} G T
\,.
\end{align}

%
%
%
%
%
%

\section{Phenomenological Implications}\label{sec:pheno}
\subsection{Flavoured Leptogenesis}\label{sec:flavour:pheno}
The full LO flavour equilibration rate has been calculated here for the first time. Adding the individual contributions, it can be expressed as
\begin{align}
	\gamma^{\rm fl} & =
\gamma^{{\rm fl}(\phi)\delta\ell}+\gamma^{{\rm fl}(\ell)\delta\ell}+
\gamma^{{\rm fl}({\rm R})\delta\ell}+\gamma^{{\rm fl}}_{{\rm vertex}}
\\\notag
&=1.32\times 10^{-3} \times  h_t^2 T +3.72\times 10^{-3} \times G T + 8.31\times 10^{-4} \times G (\log G^{-1}) T \\\notag 
	& +4.74 \times 10^{-3} \times g_1^2 T
+ 1.67\times 10^{-3} \times g_1^2 (\log g_1^{-2}) T
+ 1.7\times 10^{-3} G T \,,
\end{align}
where $G=\frac{1}{2} (3 g_2^2 + g_1^2) $. From Section~\ref{sec:fermions} we have used the numerical fit~(\ref{eqn:fermi:num}) since it is better behaved for small values of $p^0$. 

\begin{figure}[t!]
\begin{center}
\epsfig{file=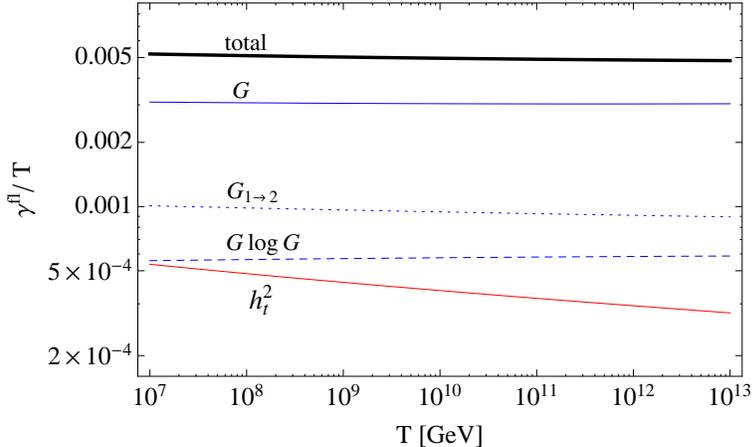,width=10cm}
\end{center}
\caption{
\label{fig:rates}
Individual contributions and total flavour equilibration rate $\gamma^{\rm fl}/T$ as function of the temperature $T$. The temperature dependence emerges through the renormalisation group equation (RGE) evolution of the coupling constants which are renormalised at the scale $2 \pi T$. Earlier times appear to the right (higher temperature) in this plot. Contributions from gauge interactions are separated into linear and log enhanced scatterings as well as $1\to 2$ processes. The $g_1^2$ and $g_1^2 (\log g_1^{-2})$ terms are included in the corresponding solid and dotted blue lines and in the total rate.  }
\end{figure}

The running of the couplings let $\gamma^{\rm fl}$ depend non-trivially on the temperature. In Figure~\ref{fig:rates} we show the individual contributions as well as the total flavour equilibration rate as function of the temperature, in the region of $T=10^7$~GeV to $10^{13}$~GeV. Note that the tree-level $1\to2$ rate is zero in this temperature regime, since the thermal masses for the Higgs and for the leptons leave no phase space for a decay process. Only once the collinear emission contributions are included a finite $\gamma^{\rm fl}_{1\leftrightarrow2}$ is obtained. 

We see that the rate is largely dominated by the terms linear in $G$, which in turn receive their dominant contribution from the $t$-channel fermion exchange, as it is shown in detail in Section~\ref{sec:fermions}. The total rate varies between $5.19\times 10^{-3}T$ and $4.83 \times 10^{-3}T$ (depending on the renormalisation scale).
This is  close to some estimates that were previously used in the literature, namely $5\times 10^{-3}T$ in~\cite{Abada:2006fw,Blanchet:2006be}, but smaller than the value of $1.75 \times 10^{-2} T$ used in~\cite{flbg}. In the light of the uncertainties due to the unknown NLO
corrections, the improvement over the popular estimate of
the flavour relaxation rate as $5\times 10^{-3}T$ may not be dramatic.
We emphasise however that in contrast to the latter number,
the relaxation rate obtained here derives from
a systematic LO calculation.
The estimates in Refs.~\cite{Abada:2006fw,Blanchet:2006be} go back
to the work~\cite{Cline:1993bd}, where the $t$-channel divergences
from fermion exchange are apparently regulated using the thermal fermion
masses, but without including their widths, such that the numerical
agreement is perhaps coincidental.
Unfortunately, the estimation of the
size of the NLO correction is not straightforward, such that
presently, we cannot state a reliable number for the theoretical
uncertainty, {\it cf.} the recently performed NLO calculation
of the photon production rate in the quark-gluon
plasma~\cite{Ghiglieri:2013gia}. If we were taking the discrepancy between
the semi-analytical expression~(\ref{gamma:t-fermi:semi-an})
for the $t$-channel scattering rate and
the numerical fit~(\ref{eqn:fermi:num})
as an indication of the theoretical uncertainty, we
would estimate it as 15\%.

When rescaling the results of Ref.~\cite{flbg} using the present LO
value for the rate of flavour relaxation, we may conclude that
the unflavoured
description of Leptogenesis 
applies to masses for the decaying right-handed neutrino
$M_N\gtrsim 10^{12}\,{\rm GeV}$, whereas
it may be treated as fully flavoured (no correlations
between the $\tau$-lepton doublets and the remaining two doublets)
below masses of $M_N\lsim 10^{11}\,{\rm GeV}$. This latter
value lies below the sometimes estimated
mass~\cite{Abada:2006ea} of
$M_N\lsim 10^{12}\,{\rm GeV}$, where a flavoured description is
assumed to be valid. Note that these are estimates obtained
for two particular points in parameter space in Ref.~\cite{flbg}, and we do not quote
a quantitative error when applying either
the unflavoured or flavoured description for a mass of the decaying
right handed neutrino between $10^{11}\,{\rm GeV}$ and
$10^{12}\,{\rm GeV}$. A more systematic and quantitative study
of the transition between the flavoured and unflavoured regimes will be
performed in the future.

The strong dependence on the gauge coupling strength also implies some model dependence of the flavour equilibration rate. In many extensions of the SM, the running of the gauge couplings is modified, such that the charged lepton Yukawa interactions will equilibrate at different temperature scales. A detailed phenomenological study of such
variants of flavoured Leptogenesis will be presented elsewhere. 
\subsection{Results for the Right-Handed Neutrino Production Rate}
\label{sec:rh:rate:results}
The integrated rate for the production of right-handed singlet neutrinos can be obtained from the results of Sections~\ref{sec:self} and~\ref{sec:vertex} by performing the $p^0$ integration with a different weight, as explained in Section~\ref{sec:rhproduction}. The main purpose of performing this integration is to illustrate the size of the corrections compared with the tree-level calculation, and to facilitate comparison with the results of Ref.~\cite{Besak:2012qm}. For a precise numerical study of Leptogenesis, one should instead use the differential distribution function $f_{N}(|\mathbf{p}|)$, since different momentum modes are equilibrated on different time scales, which can lead to a modification of the resulting lepton asymmetry~\cite{kblg}.

The contributions to the neutrino production rate from Higgs and fermion mediated scatterings, from vertex-type diagrams and from
collinearly enhanced $1\leftrightarrow 2$ processes
are respectively given by
\begin{align}
	\gamma^{N(\phi)} & = 3.15 \times 10^{-4}\times G T^4 + 5.22\times 10^{-4} \times h_t^2 T^4 \,, \label{eqn:N1prod:1}\\
	\gamma^{N(\ell)} &= 2.24 \times 10^{-3} \times G T^4 + 4.14 \times 10^{-4} \times G (\log G^{-1}) T^4 \,, \\
	\gamma^{N}_{\rm vert} & = 3.15 \times 10^{-4} \times G T^4 \,, \label{eqn:N1prod:3} \\
	\gamma^{N}_{\rm 1\to 2} & = 8.8 \times 10^{-4} \times G T^4\,,
\end{align}
where as before, $G = \frac{3}{2} g_2^2 + \frac{1}{2} g_1^2$. The rate for $\gamma^{N(\ell)}$ is calculated using the analytical decomposition into linear and logarithmic contribution analogous to the procedure
explained in Section~\ref{sec:fermions}. If instead one extracts the coefficients from a direct integration of~(\ref{C:wv:R})
and by performing a numerical fit, one obtains
\begin{align}
	\gamma^{N(\ell)} &= 1.57 \times 10^{-3} \times G T^4 + 3.67 \times 10^{-4} \times G \log G^{-1} T^4 \,.
\end{align}
See the end of Section~\ref{sec:fermions} for a detailed discussion of how these two methods and results compare.
Our numerical results
should also be compared
with those obtained in Ref.~\cite{Besak:2012qm}. While we agree with the coefficient of the terms proportional to the top quark Yukawa and of the logarithmic term, we obtain a significantly larger coefficient for the term linear in $G$. Summing the terms linear in $G$ from the contributions~(\ref{eqn:N1prod:1})-(\ref{eqn:N1prod:3}), we obtain $2.87\times 10^{-3}\times GT^4$ compared to $1.00\times 10^{-3}\times G T^4$ in Ref.~\cite{Besak:2012qm}. Using the resummed result for $\gamma^{N(\ell)}$ instead the linear coefficient is $2.20 \times 10^{-3} \times G T^4$, still deviating significantly from~Ref.~\cite{Besak:2012qm}. In
summary, our result for the production rate of right-handed neutrinos 
is
\begin{align}
\label{gammaN:total}
\gamma^N=3.08 \times 10^{-3} \times G T^4+ 3.67 \times 10^{-4} \times G \log G^{-1} T^4 + 5.22\times 10^{-4} \times h_t^2 T^4\,.
\end{align}

\begin{figure}[t!]
\begin{center}
\epsfig{file=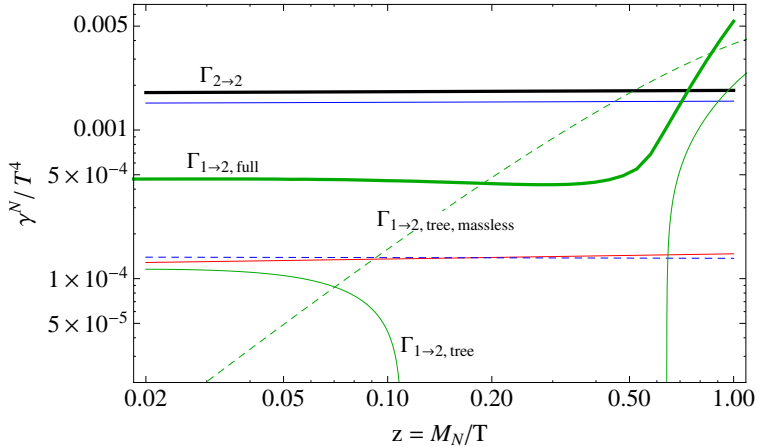,width=10cm}
\end{center}
\caption{
\label{fig:n1rates}
Individual contributions and total $N$ production rate $\gamma^{N}/T^4$, for zero $N$ density, as function of $z=M_N/T$ and $M_N = 10^{11}$~GeV. The horizontal lines show the individual and total contributions from $2\to 2$ scattering processes. 
The green solid (dashed) lines show the tree level rate for $1\to 2$ processes with (without) thermal masses for the Higgs and lepton doublet, while the thick green line shows the $1\to 2$ rate including the collinear enhancements. 
Note that as before, the $2\to 2$ scattering rates are calculated in the $M_N= 0$ approximation and therefore only valid for $z\lesssim 1$. Different from Figure~\ref{fig:rates}, time flows from left to right here as the Universe is cooling. 
}
\end{figure}
In Figure~\ref{fig:n1rates}, we show the different contributions to $N$ production for vanishing density of $N$, $f_{N}(p) = 0$. 
We see that while the collinear enhancement of $N$ production is significant when compared to the tree level rates, the gauge mediated scatterings dominate at high temperatures ($z \ll 1$), while the
tree level $\phi \to N \ell$ decay plays a negligible role for $N$ production. Before $z>1$, the $M_N =0$ approximation used for calculating the scattering rates becomes invalid. Results for the ${\cal O}(g^2)$ corrections for $N$ production in this non-relativistic
regime were recently reported in Refs.~\cite{Salvio:2011sf,Laine:2011pq}. 

Since the tree level rates are small compared to the scattering contributions in the high temperature regime, a precise calculation of the lepton asymmetry in this regime not only needs to take scatterings into account for the washout effects, but also should include them
in the calculation of the $CP$ asymmetry. This is particularly important for Leptogenesis in the weak washout regime and should
therefore be addressed in the future.

\section{Conclusions}
\label{sec:conclusions}
In this work, we have presented an approach based on the CTP formalism
to calculate interaction rates at finite temperature from 2PI
self energies. This we have employed to
calculate the flavour relaxation and the right-handed neutrino production rates relevant for Leptogenesis scenarios.
The main results of the present paper are:
\begin{itemize}
	\item We have shown that finite temperature interaction rates can be calculated perturbatively in the 2PI formalism. Using this approach, the $t$-channel divergences in diagrams with fermion exchange are automatically regulated. The linear and logarithmic dependencies on the gauge-coupling square introduced by such processes can be extracted both analytically and by a numerical fit from our calculation.
	\item The LO results for the
flavour relaxation rates and right-handed neutrino production rates and their dependence on the gauge and top Yukawa couplings are given in Section~\ref{sec:pheno}. These expressions can easily be used for obtaining interaction rates in other models, since the dependencies on the temperature, gauge-coupling evolution and hypercharge assignment are given explicitly. 
	\item We find that both rates are largely dominated by the fermion-mediated $t$-channel
scatterings, and are therefore sensitive to the RGE evolution of the gauge couplings. The coefficients vary by about 15\% depending on whether they are extracted analytically or numerically. Since the numerical fit~(\ref{eqn:fermi:num}) includes an incomplete account of higher order effects through resummation, this deviation can be taken as an indication for the magnitude of higher order effects. 
	\item While a detailed study of the impact on flavour effects remains to be done, we can already conjecture in combination with the work~\cite{flbg} that the unflavoured description of Leptogenesis will be valid for $M_N$ as low as $5\times 10^{11}$~GeV, somewhat lower than previously estimated.
\end{itemize}
The production rate of right-handed singlet neutrinos for vanishing $N$ density was previously calculated in Ref.~\cite{Besak:2012qm}. While we agree with the coefficients of the $G\log G$ and $h_t^2$ terms, we find a much larger coefficient for the contributions linear in $G$ which dominates the overall $N$ production rate. This also reduces the absolute importance of the collinear scatterings that were first shown to be relevant in~\cite{Anisimov:2010gy}. Further investigations are required to identify the origin of this discrepancy. To this end, we note that
our semi-analytical extraction of the terms linear and logarithmic
in $G$ agrees within the expected accuracy with a numerical evaluation of
the phase-space integrals, which can be taken as a consistency check.

Our calculations have been performed in the approximation that all masses of the external particles are vanishingly small. This is a good approximation for the flavour relaxation rate, where all particles have masses ${\cal O}(gT)$ that only lead to small, NLO, corrections since interactions in the plasma are dominated by hard processes with ${\cal O}(T)$ momentum exchange. For the right-handed neutrino production rate, this approximation breaks down when $M_N \gtrsim T$, i.e. at times where $z>1$. In this regime $1\to 3$ decays of $N$ are kinematically allowed and exhibit divergences when the emitted gauge boson is soft or collinear with one of the other decay products. These divergences should be cancelled by virtual corrections that arise when only two propagators are put on shell in the diagrams in Figure~ \ref{fig:fl_vertex}. For the $T\ll M_N$ limit, this cancellation was recently demonstrated in~\cite{Salvio:2011sf,Laine:2011pq}. Using the approach presented here, it has been shown that the cancellation of soft and collinear divergences
occurs for any temperature~\cite{Garbrecht:2013gd}.

Extensions and applications of the calculation of LO scattering and
production rates for light (compared to the temperature) particles
include a calculation of the $CP$-violating rate in the weak washout
regime of Leptogenesis. Besides, a systematic and quantitative study
of the transition regime between flavoured and unflavoured Leptogenesis
may now be performed in combination with the description of flavour
decoherence developed in Ref.~\cite{flbg}. The results of this
work may therefore serve as a basis for LO calculations
of the baryon asymmetry of the Universe in scenarios where so far,
only estimates have been available.

\section*{Acknowledgements}

We thank M.~Beneke for discussions and collaboration in early stages of this project. Research of PS was supported by the U.S. Department of Energy, Division of High Energy Physics, under contracts DE-AC02-06CH11357 and DE-FG02-12ER41811. BG and FG acknowledge support by the Gottfried Wilhelm Leibniz programme of the
DFG, and by the DFG cluster of excellence `Origin and Structure
of the Universe'.

\appendix

\section{RGE evolution of couplings}
The gauge and top quark Yukawa couplings at temperature scales relevant for Leptogenesis differ significantly from their values at the Electroweak scale. The RGE evolution is well known in the SM. At the
one-loop level, the gauge couplings  at a scale $\mu$ are given by 
\begin{align}
	\alpha_i^{-1}(\mu) = \alpha_i^{-1} (M_Z) - \frac{b_i}{2\pi} \log\left(\frac{\mu}{M_Z}\right)  ,
\end{align}
where $b_1 = 41/10$, $b_2 = -19/6$ and $b_3 = -7$, and the  $\alpha_i (M_Z)$ are the values of the couplings at the Electroweak scale. The couplings $\alpha_i$ are defined as $\alpha_1 = \frac{5}{3} g_1^2/(4 \pi)$ and $\alpha_{2,3} = g^2_{2,3}/(4\pi)$, following the conventions used in the context of Grand Unification. 

The RGE equations for the top Yukawa and the Higgs quartic coupling at the one loop level are given by~\cite{Lindner:1985uk} 
\begin{align}
\mu \frac{d}{d \mu} h_t^2 & =  \frac{9}{2} \frac{1}{8\pi^2} h_t^2 \left( h_t^2 - 4 \pi \left( \frac{17}{54} \alpha_1 + \frac{1}{2} \alpha_2 + \frac{16}{9} \alpha_3 \right) \right), \\
\mu \frac{d}{d \mu} \lambda & = \frac{6}{8\pi^2} \left( \lambda^2  - 4 \pi \lambda \left(\frac{\alpha_1}{4}  + \frac{3}{4} \alpha_2 - \frac{h_t^2}{4 \pi} \right) + (4 \pi)^2 \left(\frac{ \alpha_1^2}{16} + \frac{ \alpha_1 \alpha_2}{8} + \frac{3}{16} \alpha_2^2\right) - h_t^4\right),
\end{align}
where we suppress the dependence of the couplings on the RGE scale $\mu$. These are evaluated numerically using the following input values for the gauge  and top Yukawa couplings at the Electroweak scale~\cite{PDG:2011}: 
$\alpha_1 (M_Z)  = 0.0169$, 
$\alpha_2 (M_Z)  = 0.0338$, 
$\alpha_3 (M_Z)  = 0.1184$ 
and $h_t (M_Z)  = 0.998$.

The Higgs quartic coupling depends on the Higgs boson
mass through $\lambda = (m_h/v)^2/2$, where $v=174$~GeV. Current experimental constraints indicate that $m_h=125$~GeV, which we take as the value of
the Higgs boson mass for the rest of this analysis. 
\begin{figure}[t!]
\center
\epsfig{file=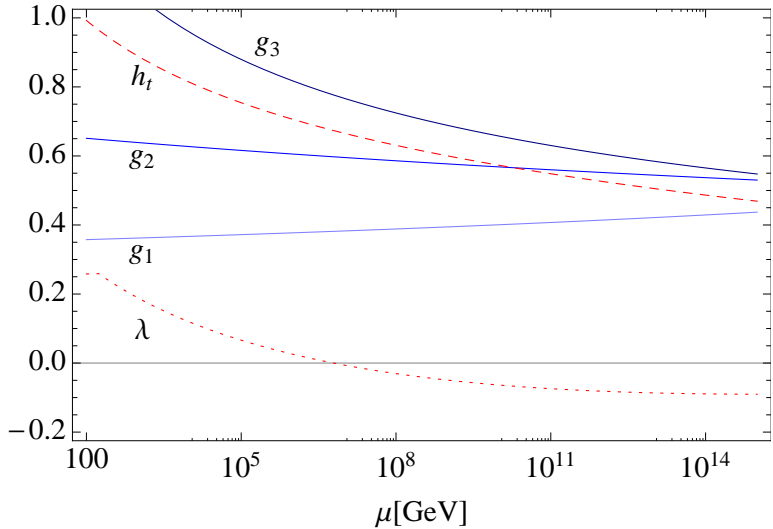}
\caption{
\label{fig:rge}
One-loop evolution of coupling constants in the SM. Note that $g_1 = \sqrt{3/5}\, g_{1,{\rm GUT}}$, such that gauge coupling unification at $\mu \sim 10^{15}$~GeV is not evident in this figure. 
}
\end{figure}

The evolution of the couplings up to scales $\mu = 10^{15}$~GeV is shown in Figure~\ref{fig:rge}. The value of the Higgs quartic coupling gets negative at intermediate scales, which could jeopardise the stability of the Electroweak vacuum. A more detailed study, including two loop corrections and threshold effects, was presented in~\cite{EliasMiro:2011aa}. They find that while the Higgs coupling indeed becomes negative at high scales, it stays above the meta-stability bound for $m_h\sim 125$~GeV, and the upper limit on the reheating temperature is consistent with the Leptogenesis scenario.
\begin{table}
\center
\begin{tabular}{|c|c|c|c|c|c|}\hline
 RGE scale & $g_1$ & $g_2$ & $g_3$ & $h_t$ & $\lambda$ \\ \hline
 $ 10^9$~GeV & 0.394 & 0.577 & 0.689 & 0.600 & -0.049 \\ \hline
 $ 10^{12}$~GeV & 0.414 & 0.552 & 0.606 & 0.526 & -0.082 \\ \hline
\end{tabular}
\caption{Values of the relevant SM coupling constants at scales important for Leptogenesis. }
\label{tab:couplings}
\end{table}

The values of the couplings for $\mu=10^9$~GeV and $\mu = 10^{12}$~GeV are given in Table~\ref{tab:couplings}. These values are used for the numerical analysis in Section~\ref{sec:flavour:pheno}. The running of the couplings in the relevant temperature regime is relatively slow, such that it is not necessary to evolve the couplings along with the temperature in a numerical analysis. The induced uncertainty is of higher order in the coupling expansion. 
%
%
%

%
%
%
%


%
%
%
\section{Feynman Rules}
For completeness, here we list the Feynman rules that are employed to calculate the one and two-loop self energies directly in Wigner space. First, note that the standard definition, in Wigner space, is such that
\begin{center}
       \begin{minipage}{.25\linewidth}
\flushright
       $ {\rm i} \Sigma\!\!\!/^{ab} (k)  =\,(-1)\times $
       \end{minipage}%
       \begin{minipage}{.25\linewidth}
	\includegraphics[width=5cm, origin=c]{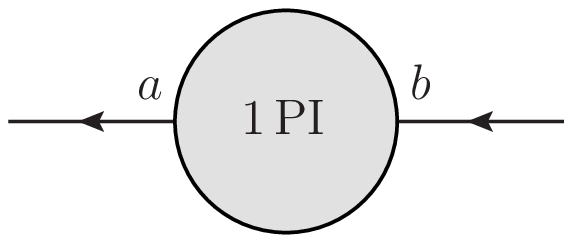}
       \end{minipage}
    \end{center}
where the blob denotes the sum of all one particle irreducible (1PI) diagrams, the momentum flows in the direction of the arrows, and the external legs are understood to be amputated.

For fermions (and similarly for complex scalars), the two-point correlation-functions on the closed time path (CTP) are defined as 
\begin{align}
	{\rm i} S(u,v) & = \left\langle T_{\cal C}( \psi(u) \bar\psi(v)) \right\rangle,
\end{align}
where $T_{\cal C}$ denotes time ordering along the CTP. Since $\bar\psi$ creates a fermion state and annihilates an anti-fermion, an arrow indicating particle flow will point from $v$ to $u$. Correspondingly, we obtain the following Feynman rules for the propagators in momentum space:
\begin{align}
    \epsfig{file=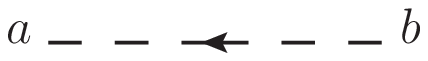,width=4cm} \!\!\!\!\!\!\!\!\!\!
     &= {\rm i} \Delta^{ab}(k)\,, \\
    \epsfig{file=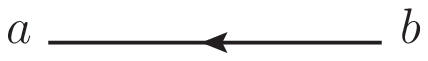,width=4cm}  \!\!\!\!\!\!\!\!\!\!
     &= {\rm i} S^{ab}(k) \,.
\end{align}
Our convention that the SU(2) doublets are defined as $\phi = ( \phi^+, \phi^0)^T$ and $\ell_i = (\nu_i, e_i^-)^T$ implies that the arrows point in the direction of positive (negative) electric
charge flow for Higgs bosons (SM leptons). The momentum flows in the direction of the arrow. Since the gauge boson propagators are neutral, the propagators have no well defined charge flow. Note that we suppress all $SU(2)$ indices in this paper. 

\paragraph{Interactions} - Interaction vertices are derived from the Lagrangian as in conventional field theory. In addition, each internal vertex obtains a sign $a \in \{+,-\}$ that is summed over. The following vertices  are relevant for the calculations in this paper: 
\begin{center}
       \begin{minipage}{.3\linewidth}
\centering
	\includegraphics[width=5cm, origin=c]{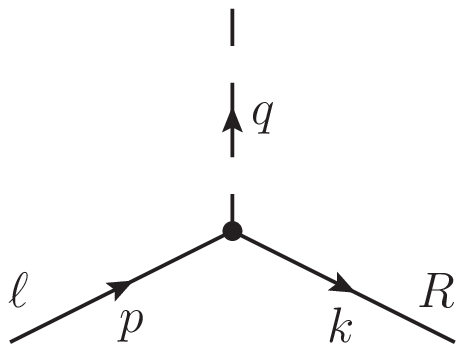}
       \end{minipage}
       \begin{minipage}{.5\linewidth}
       $\qquad  -{\rm i} h P_{\rm L}\,,$
       \end{minipage}
       \\[1cm]
        \begin{minipage}{.3\linewidth}
\centering
	\includegraphics[width=5cm, origin=c]{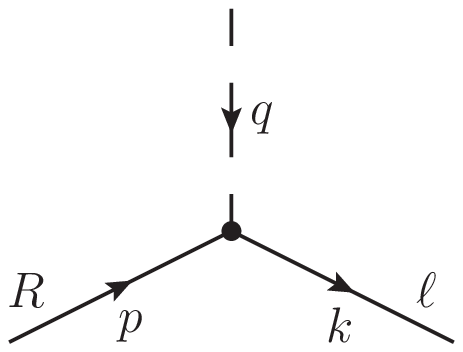}
       \end{minipage}
       \begin{minipage}{.5\linewidth}
       $\qquad  -{\rm i} h^\dagger P_{\rm R}\,,$
       \end{minipage}
\end{center}
       where we have suppressed gauge and flavour indices. For gauge boson interactions we make the gauge indices explicit:
\begin{center}
       \begin{minipage}{.3\linewidth}
\centering
	\includegraphics[width=5cm, origin=c]{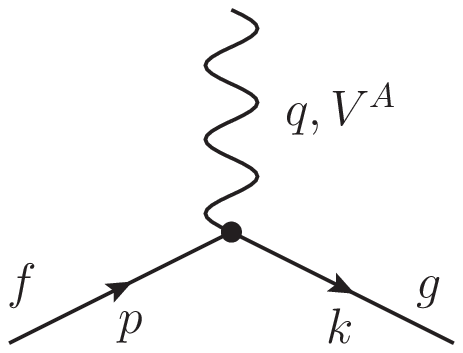}
       \end{minipage}
       \begin{minipage}{.5\linewidth}
\centering
       \begin{align} 
       &-{\rm i} g_1 \gamma^\mu (Y_{\rm L} P_{\rm L} + Y_{\rm R} P_{\rm R})\delta_{fg} \quad &&\text{for }V^A = B\,, \notag\\
       &-{\rm i} g_2 \gamma^\mu P_{\rm L} T^A_{fg} &&\text{for }V^A = W^A\,, \notag
       \end{align}
       \end{minipage}
       \\[1cm]
       \begin{minipage}{.3\linewidth}
\centering
	\includegraphics[width=5cm, origin=c]{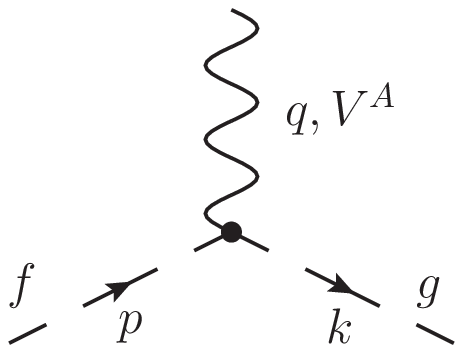}
       \end{minipage}
       \begin{minipage}{.5\linewidth}
\centering
       \begin{align} 
       &-{\rm i} g_1 Y_\phi \delta_{fg} \left( p + k\right)^\mu  \quad &&\text{for }V^A = B\,, \notag\\
       &-{\rm i} g_2 T^A_{fg} \left( p + k\right)^\mu &&\text{for }V^A = W^A\,. \notag
       \end{align}
       \end{minipage}
\end{center}
Here $T^A_{fg}$ are the generators of SU(2) in the fundamental representation, $p,k,q$ denote the momenta and $f,g$ are SU(2) indices. The scalar-scalar-vector (SSV) vertex is particularly sensitive to sign errors. The rule presented above is valid provided that both the momenta and the positive charge flows in the direction indicated by the arrows.


\begin{thebibliography}{99}

\bibitem{Fukugita:1986hr}
  M.~Fukugita and T.~Yanagida,
  ``Baryogenesis Without Grand Unification,''
  Phys.\ Lett.\ B {\bf 174} (1986) 45.
  
\bibitem{Buchmuller:2012eb}
  W.~Buchmuller,
  ``Leptogenesis: Theory and Neutrino Masses,''
  arXiv:1210.7758 [hep-ph].

\bibitem{Blanchet:2012bk}
  S.~Blanchet and P.~Di Bari,
  ``The minimal scenario of leptogenesis,''
  New J.\ Phys.\  {\bf 14} (2012) 125012
  [arXiv:1211.0512 [hep-ph]].
 
\bibitem{Fong:2013wr}
  C.~S.~Fong, E.~Nardi and A.~Riotto,
  ``Leptogenesis in the universe,''
  Adv.\ High Energy Phys.\  {\bf 2012} (2012) 158303
  [arXiv:1301.3062 [hep-ph]].


\bibitem{Schwinger:1960qe}
  J.~S.~Schwinger,
  ``Brownian motion of a quantum oscillator,''
  J.\ Math.\ Phys.\  {\bf 2} (1961) 407.

\bibitem{Keldysh:1964ud}
  L.~V.~Keldysh,
  ``Diagram technique for nonequilibrium processes,''
  Zh.\ Eksp.\ Teor.\ Fiz.\  {\bf 47} (1964) 1515
  [Sov.\ Phys.\ JETP {\bf 20} (1965) 1018].

\bibitem{Calzetta:1986cq}
  E.~Calzetta and B.~L.~Hu,
  ``Nonequilibrium Quantum Fields: Closed Time Path Effective Action, Wigner
  Function and Boltzmann Equation,''
  Phys.\ Rev.\  D {\bf 37} (1988) 2878.

\bibitem{Buchmuller:2000nd} 
  W.~Buchmuller and S.~Fredenhagen,
  ``Quantum mechanics of baryogenesis,''
  Phys.\ Lett.\ B {\bf 483}, 217 (2000)
  [hep-ph/0004145].


\bibitem{De Simone:2007rw}
  A.~De Simone and A.~Riotto,
  ``Quantum Boltzmann Equations and Leptogenesis,''
  JCAP {\bf 0708} (2007) 002
  [hep-ph/0703175].


\bibitem{Cirigliano:2009yt}
  V.~Cirigliano, C.~Lee, M.~J.~Ramsey-Musolf and S.~Tulin,
  ``Flavored Quantum Boltzmann Equations,''
  Phys.\ Rev.\ D {\bf 81} (2010) 103503
  [arXiv:0912.3523 [hep-ph]].


\bibitem{flbg}
  M.~Beneke, B.~Garbrecht, C.~Fidler, M.~Herranen and P.~Schwaller,
  ``Flavoured Leptogenesis in the CTP Formalism,''
  Nucl.\ Phys.\ B {\bf 843} (2011) 177
  [arXiv:1007.4783 [hep-ph]].
  
\bibitem{Anisimov:2010dk}
  A.~Anisimov, W.~Buchmuller, M.~Drewes and S.~Mendizabal,
  ``Quantum Leptogenesis I,''
  Annals Phys.\  {\bf 326} (2011) 1998
  [arXiv:1012.5821 [hep-ph]].


\bibitem{Covi:1997dr}
  L.~Covi, N.~Rius, E.~Roulet and F.~Vissani,
  ``Finite temperature effects on CP violating asymmetries,''
  Phys.\ Rev.\ D {\bf 57} (1998) 93
  [hep-ph/9704366].

\bibitem{Giudice:2003jh}
  G.~F.~Giudice, A.~Notari, M.~Raidal, A.~Riotto and A.~Strumia,
  ``Towards a complete theory of thermal leptogenesis in the SM and MSSM,''
  Nucl.\ Phys.\ B {\bf 685} (2004) 89
  [hep-ph/0310123].

\bibitem{Garny:2009rv}
  M.~Garny, A.~Hohenegger, A.~Kartavtsev and M.~Lindner,
  ``Systematic approach to leptogenesis in nonequilibrium QFT: vertex
  contribution to the CP-violating parameter,''
  Phys.\ Rev.\  D {\bf 80} (2009) 125027
  [arXiv:0909.1559 [hep-ph]].

\bibitem{Garny:2009qn}
  M.~Garny, A.~Hohenegger, A.~Kartavtsev and M.~Lindner,
  ``Systematic approach to leptogenesis in nonequilibrium QFT: self-energy
  contribution to the CP-violating parameter,''
  Phys.\ Rev.\  D {\bf 81} (2010) 085027
  [arXiv:0911.4122 [hep-ph]].

\bibitem{Anisimov:2010aq}
  A.~Anisimov, W.~Buchm\"uller, M.~Drewes and S.~Mendizabal,
  ``Leptogenesis from Quantum Interference in a Thermal Bath,''
  Phys.\ Rev.\ Lett.\  {\bf 104} (2010) 121102
  [arXiv:1001.3856 [hep-ph]].

\bibitem{Garny:2010nj}
  M.~Garny, A.~Hohenegger, A.~Kartavtsev,
  ``Medium corrections to the CP-violating parameter in leptogenesis,''
  Phys.\ Rev.\  {\bf D81 } (2010)  085028.
  [arXiv:1002.0331 [hep-ph]].

\bibitem{kblg}
  M.~Beneke, B.~Garbrecht, M.~Herranen and P.~Schwaller,
  ``Finite Number Density Corrections to Leptogenesis,''
  Nucl.\ Phys.\ B {\bf 838} (2010) 1
  [arXiv:1002.1326 [hep-ph]].
  
  
\bibitem{Garbrecht:2010sz}
  B.~Garbrecht,
  ``Leptogenesis: The Other Cuts,''
  Nucl.\ Phys.\  {\bf B847 } (2011)  350-366.
  [arXiv:1011.3122 [hep-ph]].
  
\bibitem{Garbrecht:2011aw}
  B.~Garbrecht and M.~Herranen,
  ``Effective Theory of Resonant Leptogenesis in the Closed-Time-Path Approach,''
  Nucl.\ Phys.\ B {\bf 861} (2012), 17.
  [arXiv:1112.5954 [hep-ph]].

\bibitem{Garny:2011hg}
  M.~Garny, A.~Kartavtsev and A.~Hohenegger,
  ``Leptogenesis from first principles in the resonant regime,''
  Annals Phys.\  {\bf 328} (2013) 26
  [arXiv:1112.6428 [hep-ph]].

\bibitem{Garbrecht:2012qv}
  B.~Garbrecht,
  ``Leptogenesis from Additional Higgs Doublets,''
  Phys.\ Rev.\ D {\bf 85} (2012) 123509
  [arXiv:1201.5126 [hep-ph]].

\bibitem{Drewes:2012ma}
  M.~Drewes and B.~Garbrecht,
  ``Leptogenesis from a GeV Seesaw without Mass Degeneracy,''
  to appear in JHEP
  [arXiv:1206.5537 [hep-ph]].


\bibitem{Garbrecht:2012pq}
  B.~Garbrecht,
  ``Baryogenesis from Mixing of Lepton Doublets,''
  Nucl.\ Phys.\ B {\bf 868} (2013) 557
  [arXiv:1210.0553 [hep-ph]].

  
\bibitem{Frossard:2012pc}
  T.~Frossard, M.~Garny, A.~Hohenegger, A.~Kartavtsev and D.~Mitrouskas,
  ``Systematic approach to thermal leptogenesis,''
  arXiv:1211.2140 [hep-ph].
  
\bibitem{Millington:2012pf}
  P.~Millington and A.~Pilaftsis,
  ``Perturbative Non-Equilibrium Thermal Field Theory,''
  arXiv:1211.3152 [hep-ph].

\bibitem{Endoh:2003mz} 
  T.~Endoh, T.~Morozumi and Z.~-h.~Xiong,
  ``Primordial lepton family asymmetries in seesaw model,''
  Prog.\ Theor.\ Phys.\  {\bf 111}, 123 (2004)
  [hep-ph/0308276].
\bibitem{Abada:2006fw} 
  A.~Abada, S.~Davidson, F.~-X.~Josse-Michaux, M.~Losada and A.~Riotto,
  ``Flavor issues in leptogenesis,''
  JCAP {\bf 0604}, 004 (2006)
  [hep-ph/0601083].
\bibitem{Nardi:2006fx} 
  E.~Nardi, Y.~Nir, E.~Roulet and J.~Racker,
  ``The Importance of flavor in leptogenesis,''
  JHEP {\bf 0601}, 164 (2006)
  [hep-ph/0601084].


\bibitem{Anisimov:2010gy}
  A.~Anisimov, D.~Besak, D.~Bodeker,
  ``Thermal production of relativistic Majorana neutrinos: Strong enhancement by multiple soft scattering,''
  JCAP {\bf 1103 } (2011)  042.
  [arXiv:1012.3784 [hep-ph]].
  
\bibitem{Besak:2012qm} 
  D.~Besak and D.~Bodeker,
  ``Thermal production of ultrarelativistic right-handed neutrinos: Complete leading-order results,''
  JCAP {\bf 1203}, 029 (2012)
  [arXiv:1202.1288 [hep-ph]].

\bibitem{Kiessig:2010pr}
  C.~P.~Kiessig, M.~Plumacher and M.~H.~Thoma,
  ``Decay of a Yukawa fermion at finite temperature and applications to leptogenesis,''
  Phys.\ Rev.\ D {\bf 82} (2010) 036007
  [arXiv:1003.3016 [hep-ph]].

\bibitem{Kiessig:2011fw}
  C.~Kiessig and M.~Plumacher,
  ``Hard-Thermal-Loop Corrections in Leptogenesis I: CP-Asymmetries,''
  JCAP {\bf 1207} (2012) 014
  [arXiv:1111.1231 [hep-ph]].

\bibitem{Kiessig:2011ga}
  C.~Kiessig and M.~Plumacher,
  ``Hard-Thermal-Loop Corrections in Leptogenesis II: Solving the Boltzmann Equations,''
  JCAP {\bf 1209} (2012) 012
  [arXiv:1111.1235 [hep-ph]].


\bibitem{Arnold:2000dr}
  P.~B.~Arnold, G.~D.~Moore and L.~G.~Yaffe,
  ``Transport coefficients in high temperature gauge theories. 1. Leading log results,''
  JHEP {\bf 0011} (2000) 001
  [hep-ph/0010177].

\bibitem{Arnold:2001ms}
  P.~B.~Arnold, G.~D.~Moore and L.~G.~Yaffe,
  ``Photon emission from quark gluon plasma: Complete leading order results,''
  JHEP {\bf 0112} (2001) 009
  [hep-ph/0111107].

\bibitem{Salvio:2011sf}
  A.~Salvio, P.~Lodone and A.~Strumia,
  ``Towards leptogenesis at NLO: the right-handed neutrino interaction rate,''
  JHEP {\bf 1108} (2011) 116
  [arXiv:1106.2814 [hep-ph]].

\bibitem{Laine:2011pq}
  M.~Laine and Y.~Schroder,
  ``Thermal right-handed neutrino production rate in the non-relativistic regime,''
  JHEP {\bf 1202} (2012) 068
  [arXiv:1112.1205 [hep-ph]].

\bibitem{Garbrecht:2013gd}
  B.~Garbrecht, F.~Glowna and M.~Herranen,
  ``Right-Handed Neutrino Production at Finite Temperature: Radiative Corrections, Soft and Collinear Divergences,''
 to appear in JHEP
  [arXiv:1302.0743 [hep-ph]].
  
\bibitem{Prokopec:2003pj}
  T.~Prokopec, M.~G.~Schmidt and S.~Weinstock,
  ``Transport equations for chiral fermions to order h bar and electroweak baryogenesis. Part 1,''
  Annals Phys.\  {\bf 314} (2004) 208
  [hep-ph/0312110].
  
\bibitem{Garbrecht:2008cb} 
  B.~Garbrecht and T.~Konstandin,
  ``Separation of Equilibration Time-Scales in the Gradient Expansion,''
  Phys.\ Rev.\ D {\bf 79}, 085003 (2009)
  [arXiv:0810.4016 [hep-ph]].

\bibitem{Ghiglieri:2013gia}
  J.~Ghiglieri, J.~Hong, A.~Kurkela, E.~Lu, G.~D.~Moore and D.~Teaney,
  ``Next-to-leading order thermal photon production in a weakly coupled quark-gluon plasma,''
  arXiv:1302.5970 [hep-ph].

\bibitem{Blanchet:2006be}
  S.~Blanchet and P.~Di Bari,
  ``Flavor effects on leptogenesis predictions,''
  JCAP {\bf 0703} (2007) 018
  [hep-ph/0607330].

\bibitem{Abada:2006ea}
  A.~Abada, S.~Davidson, A.~Ibarra, F.~-X.~Josse-Michaux, M.~Losada and A.~Riotto,
  ``Flavour Matters in Leptogenesis,''
  JHEP {\bf 0609} (2006) 010
  [hep-ph/0605281].

\bibitem{Cline:1993bd}
  J.~M.~Cline, K.~Kainulainen and K.~A.~Olive,
  ``Protecting the primordial baryon asymmetry from erasure by sphalerons,''
  Phys.\ Rev.\ D {\bf 49} (1994) 6394
  [hep-ph/9401208].

\bibitem{Lindner:1985uk}
  M.~Lindner,
  ``Implications of Triviality for the Standard Model,''
  Z.\ Phys.\ C {\bf 31} (1986) 295.
  
\bibitem{PDG:2011}
  K.~Nakamura {\it et al.}  [Particle Data Group Collaboration],
  ``Review of particle physics,''
  J.\ Phys.\ G G {\bf 37} (2010) 075021.
  
\bibitem{EliasMiro:2011aa} 
  J.~Elias-Miro, J.~R.~Espinosa, G.~F.~Giudice, G.~Isidori, A.~Riotto and A.~Strumia,
  ``Higgs mass implications on the stability of the electroweak vacuum,''
  arXiv:1112.3022 [hep-ph].


\end{thebibliography}
\end{document}